\documentclass[12pt]{article}

\usepackage{graphicx}
\usepackage{fancybox}
\usepackage{amsmath}
\usepackage{amssymb}
\usepackage{latexsym}
\usepackage{epsfig}

\newcommand{\al}{\alpha}
\newcommand{\ep}{\epsilon}
\newcommand{\vep}{\varepsilon}

\newcommand{\NS}{\mbox{NS}}
\newcommand{\tNS}{\widetilde{\mbox{NS}}}
\newcommand{\R}{\mbox{R}}
\newcommand{\tR}{\widetilde{\mbox{R}}}
\newcommand{\sNS}{\msc{NS}}
\newcommand{\stNS}{\widetilde{\msc{NS}}}
\newcommand{\sR}{\msc{R}}
\newcommand{\stR}{\widetilde{\msc{R}}}

\newcommand{\lb}{\lbrack}
\newcommand{\rb}{\rbrack}

\newcommand{\msc}[1]{\mbox{\scriptsize #1}}
\newcommand{\dsp}{\displaystyle}

\newcommand{\bc}{\Bbb C}
\newcommand{\br}{\Bbb R}
\newcommand{\bz}{\Bbb Z}

\newcommand{\bh}{{\Bbb H}}

\newcommand{\bsz}{\Bbb Z}

\newcommand{\cJ}{{\cal J}}

\newcommand{\cO}{{\cal O}}
\newcommand{\cN}{{\cal N}}

\newcommand{\cH}{{\cal H}}

\newcommand{\cK}{{\cal K}}
\newcommand{\cZ}{{\cal Z}}

\newcommand{\tL}{\tilde{L}}
\newcommand{\tJ}{\tilde{J}}

\newcommand{\tell}{\tilde{\ell}}

\newcommand{\tZ}{\widetilde{Z}}
\newcommand{\tcZ}{\widetilde{\cZ}}

\newcommand{\tz}{\widetilde{z}}

\newcommand{\hc}{\hat{c}}

\newcommand{\hcK}{\widehat{\cal K}}

\newcommand{\Th}[2]{\Theta_{#1,#2}}
\renewcommand{\th}{{\theta}}

\newcommand{\ch}[2]{\mbox{ch}^{#1}_{#2}}

\newcommand{\chic}{\chi_{\msc{\bf con}}}

\newcommand{\chid}{\chi_{\msc{\bf dis}}}

\newcommand{\hchid}{\widehat{\chi}_{\msc{\bf dis}}}

\newcommand{\tr}{\mbox{Tr}}

\newcommand{\nn}{\nonumber\\}

\newcommand{\reg}{\bf reg}

\newcommand{\any}{{}^{\forall}}

\renewcommand{\mod}{\, \mbox{mod} ~ }

\def\boxit#1{\vbox{\hrule\hbox{\vrule\kern8pt
\vbox{\hbox{\kern8pt}\hbox{\vbox{#1}}\hbox{\kern8pt}}
\kern8pt\vrule}\hrule}}
\def\mathboxit#1{\vbox{\hrule\hbox{\vrule\kern8pt\vbox{\kern8pt
\hbox{$\displaystyle #1$}\kern8pt}\kern8pt\vrule}\hrule}}

\newcommand {\eqn}[1]{(\ref{#1})}

\makeatletter
\@addtoreset{equation}{section}
\def\theequation{\thesection.\arabic{equation}}
\makeatother

\begin{document}

\begin{titlepage}
 \
 \renewcommand{\thefootnote}{\fnsymbol{footnote}}
 \font\csc=cmcsc10 scaled\magstep1
 {\baselineskip=14pt
 \rightline{
 \vbox{\hbox{November, 2013}
       }}}

 \baselineskip=20pt
 
\begin{center}

{\bf \Large  `Analytic Continuation' of $\cN=2$ Minimal Model

} 

 
\vskip 2cm
 
\noindent{ \large Yuji Sugawara}\footnote{\sf ysugawa@se.ritsumei.ac.jp}
\\

\medskip

 {\it Department of Physical Sciences, 
 College of Science and Engineering, \\ 
Ritsumeikan University,  
Shiga 525-8577, Japan}
 

\end{center}

\bigskip

\begin{abstract}

In this paper we discuss what theory should be identified as the `analytic continuation' with $N \rightarrow -N$
of the ${\cal N}=2$ minimal model with the central charge $\hat{c} = 1 - \frac{2}{N}$.
We clarify how the elliptic genus of the expected model is written in terms of  {\em holomorphic\/} linear combinations of 
the `modular completions' introduced in \cite{ES-NH} 
in the $SL(2)_{N+2}/U(1)$-supercoset theory.
We further discuss how this model could be interpreted as a kind of {\em `compactified'} model of the $SL(2)_{N+2}/U(1)$-supercoset 
in the $(\tR,\tR)$-sector, in which only the discrete spectrum appears in the torus partition function and 
the potential IR-divergence due to the non-compactness of target space is removed. 
We also briefly argue on possible definitions of the sectors with other spin structures.

\end{abstract}


\setcounter{footnote}{0}
\renewcommand{\thefootnote}{\arabic{footnote}}

\end{titlepage}

\baselineskip 18pt

\vskip2cm 


\section{Introduction}

The $\cN=2$ minimal model is one of the most familiar rational superconformal field theories in two-dimension \cite{N=2 minimal}. 
This is defined by the supercoset theory of $SU(2)_k/U(1)$ with level $k= N-2$ \cite{KS} and also described as the IR-fixed point of the $\cN=2$
Landau-Ginzburg (LG) model with the superpotential $W(X) = X^{N} + [\mbox{lower powers}]$ \cite{N=2LG}. 
One of  good features of $\cN=2$ minimal models is a very simple formula of elliptic genus \cite{Witten-EG minimal} ;   
\begin{equation}
\cZ^{(N-2)}_{\msc{min}}(\tau,z) = \frac{\th_1\left(\tau, \frac{N-1}{N}z\right)}
{\th_1\left(\tau, \frac{1}{N}z\right)},
\label{Z min}
\end{equation}
which nicely behaves under modular transformations as well as the spectral flows. 
Namely, the function $\cZ^{(N-2)}_{\msc{min}}(\tau,Nz)$ is 
a weak Jacobi form  \cite{EZ} of weight 0 and 
index $\frac{N(N-2)}{2}$. 

It is an old question {\em what theory should be identified as the `analytic continuation' of the $\cN=2$ minimal model under $N\, \rightarrow \, -N$.} 
A naive guess of the answer would be the $SL(2)_{N+2}/U(1)$-supercoset theory that has the expected central charge 
$\hc \left( \equiv \frac{c}{3} \right) = 1 + \frac{2}{N} $. 
However, it is {\em not} completely correct  due to the following reasons;
\begin{itemize}
\item The $SL(2)/U(1)$-supercoset theory contains both discrete and continuous spectra of primary fields, while  
the $\cN=2$ minimal model only has discrete spectra. It is not likely to be the case that such two theories are directly connected 
by an analytic continuation of the parameter of theory. 


\item It has been known that the elliptic genera of the $SL(2)/U(1)$-supercoset theories shows non-holomorphicity with respect to 
the modulus $\tau$ of the world-sheet torus  \cite{Troost,ES-NH}, whereas 
the elliptic genera of the minimal model \eqn{Z min}
are manifestly holomorphic.

\end{itemize}


In this paper we shall try to give a precise answer to this question. 
In other words,
we will focus on  the problems;
\begin{description}
\item[(i)] 
What is the superconformal model with $\hc = 1 + \frac{2}{N}$ which has 
\begin{equation}
\cZ(\tau,z) (\propto "\cZ_{\msc{min}}^{(-N-2)}(\tau,z)")
 = \frac{\th_1\left(\tau, \frac{N+1}{N}z\right)}
{\th_1\left(\tau, \frac{1}{N}z\right)},
\label{Z expected}
\end{equation}
as its elliptic genus?

\item[(ii)]
Then, does this  model  have any relationship with the $SL(2)_{N+2}/U(1)$-supercoset?

\end{description}

~


\newpage

This paper is organized as follows:

In section 2 we shall demonstrate our mathematical results. We make a detailed analysis   
on the holomorphic function \eqn{Z expected}, and prove
the main theorem which addresses the precise relation 
between \eqn{Z expected} and the `modular completions' introduced in \cite{ES-NH}
in the $SL(2)_{N+2}/U(1)$-supercoset. 
We will further present a physical interpretation of this mathematical result 
and some discussions in section 3. 
In section 4 we present the summary and some comments.

~


\section{Holomorphic Linear Combinations of Modular Completions }

In this section we address some mathematical results. 
The main claims will be expressed in \eqn{X hchid} and \eqn{Z tZ Phi}.

~


\subsection{Preliminary}

We first prepare relevant notations. 
To begin with, we introduce the symbol of `IR-part' just for convenience;
\begin{equation}
\left[f(\tau,z) \right] := \lim_{\tau \, \rightarrow\, i \infty}\, f(\tau,z),
\label{def IR part}
\end{equation}
where $f(\tau,z)$ is assumed to be holomorphic around the cusp $\tau= i\infty$.

~


\subsubsection{A Holomorphic Jacobi Form}

We consider a holomorphic function defined by 
\begin{equation}
\Phi^{(N)}(\tau,z) : = 
\frac{\th_1\left(\tau, \frac{N+1}{N}z\right)}
{\th_1\left(\tau, \frac{1}{N}z\right)}.
\label{Phi N}
\end{equation}
This is  obtained by the formal replacement  $N \rightarrow -N$ in 
the elliptic genus of $\cN=2$ minimal model 
\eqn{Z min}.
The function \eqn{Phi N} possesses the next modular and spectral flow 
properties with $ \hc \equiv 1 + \frac{2}{N}$;
\begin{eqnarray}
\Phi^{(N)}(\tau+1, z) =\Phi^{(N)}(\tau,z), \hspace{1cm}
\Phi^{(N)} \left(-\frac{1}{\tau}, \frac{z}{\tau} \right)
= e^{i\pi \frac{\hat{c}}{\tau} z^2}\, \Phi^{(N)} (\tau,z). 
\label{modular Phi N}
\end{eqnarray}
\begin{equation}
\Phi^{(N)}(\tau,z+m\tau+n) = (-1)^{m+n} q^{-\frac{\hc}{2}m^2}
y^{-\hc m} \, \Phi^{(N)} (\tau,z),
\hspace{1cm} (m,n\in N\bz).
\label{sflow Phi N}
\end{equation}
In other words, $\Phi^{(N)}(\tau, Nz)$ is a weak Jacobi form of weight 0 and index $\frac{N^2 \hc}{2} \equiv \frac{N(N+2)}{2}$;
$$
\Phi^{(N)}(\tau, Nz) \in \cJ[0, \frac{N(N+2)}{2}],
$$
where we set
\begin{equation}
\cJ [w,d] := \left\{\mbox{weak Jacobi forms of weight $w$ and index $d$}\right\}.
\label{set of Jacobi form}
\end{equation}

The  IR-part of \eqn{Phi N} is evaluated as
\begin{equation}
\left[
\Phi^{(N)}(\tau,z) \right]
= y^{-\frac{1}{2}} \sum_{v=0}^{N} y^{\frac{v}{N}}.
\label{IR Phi}
\end{equation}


We next introduce the `spectral flow operator' $s_{(a,b)}$ ($a,b\in \bz$)
defined by
\begin{equation}
s_{(a,b)} \cdot f(\tau,z) := (-1)^{a+b} q^{\frac{\hc}{2} a^2} y^{\hc a}
e^{2\pi i \frac{ab}{N}}\, f(\tau,z+a\tau+b),
\label{def s ab}
\end{equation}
and set
\begin{eqnarray}
\Phi^{(N)}_{(a,b)}(\tau,z) &:= & 
s_{(a,b)}\cdot \Phi^{(N)}(\tau,z)
\hspace{1cm}
(a,b\in \bz)
\label{Phi ab}
\end{eqnarray}
Since having a periodicity 
$$
\Phi_{(a+Nm,b+Nn)}^{(N)} (\tau,z) =  \Phi^{(N)}_{(a,b)} (\tau,z),
\hspace{1cm}
(\any m,n \in \bz),
$$
we may assume $a, b\in \bz_N$.
Its modular property is written as  
\begin{equation}
\Phi^{(N)}_{(a,b)}(\tau+1, z) =\Phi_{(a,a+b)}^{(N)}(\tau,z), 
\hspace{1cm}
\Phi_{(a,b)}^{(N)} \left(-\frac{1}{\tau}, \frac{z}{\tau} \right)
= e^{i\pi \frac{\hat{c}}{\tau} z^2}\, \Phi_{(b,-a)}^{(N)} (\tau,z), 
\label{modular Phi ab}
\end{equation}
The IR-part of 
$\Phi^{(N)}_{(a,b)}(\tau,z)$ is computed as 
\begin{eqnarray}
\left[\Phi^{(N)}_{(a,b)}(\tau,z) \right]
&=& \delta_{a,0}^{(N)} \,  \sum_{v=1}^{N} \, e^{2\pi i \frac{b}{N}v}
y^{-\frac{1}{2} + \frac{v}{N}}
+   y^{- \frac{1}{2} +\frac{[a]}{N}},
\label{IR Phi ab}
\end{eqnarray}
where we introduced the notation $[a]$, defined by 
$[a] \equiv a ~ (\mod\, N)$, $0\leq [a] \leq N-1$,
and $\delta^{(N)}_{r,s}$ denotes the `mod $N$ Kronecker delta'.  
In fact, we find 
$$
\Phi^{(N)}_{(0,b)}(\tau,z) = 
\frac{y^{-\frac{N+1}{2N}} e^{2\pi i \frac{b}{2N}} 
\left(1- y^{\frac{N+1}{N}} e^{2\pi i \frac{b}{N}}  \right)}
{y^{-\frac{1}{2N}} e^{2\pi i \frac{b}{2N}} 
\left(1- y^{\frac{1}{N}}e^{2\pi i \frac{b}{N}} \right)}
+ O(q)
= \sum_{v=0}^N\, e^{2\pi i \frac{b}{N}v}y^{-\frac{1}{2}+ \frac{v}{N}} + O(q),
$$
and also (for $a\neq 0$), 
$$
\Phi^{(N)}_{(a,b)}(\tau,z) 
= y^{-\frac{1}{2}+ \frac{[a]}{N}} + O(q^{\frac{1}{N}}),
$$
which proves \eqn{IR Phi ab}.



~


\subsubsection{Modular Completions}

Let us  introduce the `modular completion'  
of the extended discrete characters  (of the $\tR$-sector)   \cite{ES-L,ES-BH,ES-C}
in the $SL(2)/U(1)$-supercoset according to \cite{ES-NH}.
For the case of $\hc = 1 + \frac{2}{N}, ~ (\any N \in \bz_{>0}) $,
this function is defined as 
\begin{eqnarray}
\hspace{-1cm}
\hchid^{(N,1)} (v,a;\tau,z) 
&: =& 
\frac{\th_1(\tau,z)}{ 2\pi \eta(\tau)^3}\,
\sum_{\stackrel{n\in a + N\bz}{r \in v + N\bz}}\, 
\left\{ \int_{\br + i(N-0)} dp\, -
\int_{\br-i0} dp \, \left(y q^{n} \right) 
\right\}
\,
\frac{ e^{- \pi \tau_2 \frac{p^2+r^2}{N}} 
\left(y q^{n}\right)^{\frac{r}{N}}}{p-ir} 
\, \frac{y^{\frac{2n}{N}} q^{\frac{n^2}{N}}}{1-y q^{n}}
\nn
&\equiv & \chid^{(N,1)}(v,a;\tau,z) 
+ \frac{\th_1(\tau,z)}{ 2\pi \eta(\tau)^3}\,
\sum_{\stackrel{n\in a + N\bz}{r \in v + N\bz}}\, 
\int_{\br -i 0} dp \,
\frac{ e^{- \pi \tau_2 \frac{p^2+r^2}{N}} }{p-ir} 
\, \left(y q^{n}\right)^{\frac{r}{N}} y^{\frac{2n}{N}} q^{\frac{n^2}{N}}.
\nn
&&
\hspace{9cm} (v, a\in \bz_N) ,
\label{hchid K=1}
\end{eqnarray}
where $\chid^{(N,1)}$ denotes the extended discrete character introduced in \cite{ES-L,ES-BH,ES-C} (written in the convention of \cite{ES-NH});
\begin{eqnarray}
\chid^{(N,1)} (v,a;\tau,z) &\equiv  & 
\sum_{n\in a + N\bz}\,  \frac{(yq^{n})^{\frac{v}{N}}}
{1-yq^{n}} \, y^{\frac{2n}{N}}q^{\frac{n^2}{N} }\, \frac{\th_1(\tau,z)}{i \eta(\tau)^3}\,,
\nn
&& 
\hspace{2cm} (v=0,1,\ldots, N, ~~ a\in \bz_N) .
\label{chid K=1}
\end{eqnarray}
The first line \eqn{hchid K=1} naturally appears through  the analysis of partition function of 
$SL(2)/U(1)$-supercoset \cite{ES-NH}, and the second line just  comes from the contour deformation.


The modular completion of Appell function (or the `Appell-Lerch sum) $\cK^{(2N)}(\tau,z)$ \cite{Pol,STT} is 
given as \cite{Zwegers}\footnote
   {The relation to  the notation given in  Chapter 3 of \cite{Zwegers} are as follows; 
$$
\cK^{(2k)}(\tau,z) \equiv f^{(k)}_{u=0}(z;\tau), \hspace{1cm}
\hcK^{(2k)}(\tau,z) \equiv \tilde{f}^{(k)}_{u=0} (z;\tau).
$$
}; 
\begin{eqnarray}
\hcK^{(2N)} (\tau, z) &:=& \cK^{(2N)}(\tau,z) - \frac{1}{2} \sum_{m\in \bz_{2N}}\,
R_{m,N}(\tau) \, \Th{m}{N}(\tau,2z),
\label{hcK mt}
\end{eqnarray}
where we set 
\begin{eqnarray}
R_{m,N}(\tau) &:=& 
\frac{1}{i\pi}\, \sum_{r \in m+2N\bz}\,
\int_{\br- i0} dp \, \frac{e^{-\pi \tau_2 \frac{p^2+r^2}{N}} }{p-ir}\,
q^{- \frac{r^2}{4N}},
\label{RmN mt}
\end{eqnarray}
which is generically non-holomorphic due to explicit $\tau_2$-dependence.



The next `Fourier expansion relation' \cite{ES-NH} will be useful for our analysis; 
\begin{eqnarray}
&& y^{\frac{2a}{N} } q^{\frac{a^2}{N} } \,
 \hcK^{(2N)}\left(\tau, \frac{z+a\tau+b}{N}\right)\, 
\frac{\th_1(\tau,z)}{i \eta(\tau)^3} 
= \sum_{v\in \bz_N} \,  e^{2\pi i \frac{v b}{N}} \, \hchid^{(N,1)} (v,a;\tau,z),
\label{rel hchid hcK K-1} 
\end{eqnarray}
which is the `hatted' version of the similar relation between $\cK^{(2k)}$ and $\chid(v,a)$ 
given in \cite{ES-C}.


The modular transformation formulas for the modular completions
\eqn{hchid K=1}, \eqn{hcK mt} are written as \cite{ES-NH,Zwegers}
\begin{eqnarray}
 \hspace{-1cm}
\hchid^{(N,1)} \left(v,a ; - \frac{1}{\tau}, \frac{z}{\tau}\right)
&=& e^{i\pi \frac{\hc}{\tau}z^2}\, \sum_{v'=0}^{N-1} \,\sum_{a'\in \bz_N}\,
\frac{1}{N} \, e^{2\pi i \frac{vv' - (v+2a)(v'+2a')}{2N}}
\, \hchid^{(N,1)} (v',a';\tau,z),
\label{S hchid K=1}
\\
\hchid^{(N,1)} \left(v,a ; \tau+1, z \right)
&=& e^{2\pi i \frac{a}{N} \left(v+ a \right)}\,
\hchid^{(N,1)} \left(v,a ; \tau, z \right).
\label{T hchid K=1}
\\
\hcK^{(2N)} \left(-\frac{1}{\tau}, \frac{z}{\tau} \right) &=& \tau e^{i\pi \frac{2N}{\tau}z^2}\, \hcK^{(2N)}(\tau,z), 
\label{S hcK mt}
\\
\hcK^{(2N)} (\tau+1, z) &=& \hcK^{(2N)} (\tau, z).
\label{T hck mt}
\end{eqnarray}
The spectral flow property is summarized as follows;
\begin{eqnarray}
\hchid^{(N,1)} (v,a;\tau,z+r\tau+s) &=& (-1)^{r+s} e^{2\pi i \frac{v+2a}{N}s}
q^{-\frac{\hc}{2}r^2} y^{-\hc r}\, \hchid^{(N,1)}(v,a+r;\tau,z),
\nn
&& \hspace{7cm} (\any r,s \in \bz).
\label{sflow hchid K=1}
\\
\hcK^{(2N)} (\tau, z+ r\tau+s) &=& q^{-N r^2} y^{-2N r} \hcK^{(2N)}(\tau,z),
~~~ (\any r,s \in \bz).
\label{sflow hcK mt}
\end{eqnarray}

~


More detailed formulas 
in general cases of $\hchid^{(N,K)}(v,a)$ with 
 $\hc = 1 + \frac{2K}{N}$, $(N,K \in \bz_{>0})$ are summarized in Appendix B.

~


\subsection{Fourier Expansion of \eqn{Phi N} and Modular Completions}

Let us start our main analysis. 
We begin with  introducing  the holomorphic functions $X^{(N)}(v,a;\tau,z)$ as the `Fourier 
transforms' of $\Phi_{(a,b)}^{(N)}(\tau,z)$; 
\begin{equation}
X^{(N)}(v,a;\tau,z) := \frac{1}{N} \sum_{b\in \bz_N}\, 
e^{-2\pi i \frac{b}{N}(v+a)}\, \Phi_{(a,b)}^{(N)} (\tau,z),
\label{Xva def}
\end{equation}
in other words, 
\begin{equation}
\Phi_{(a,b)}^{(N)} (\tau,z) = \sum_{v=0}^{N-1} \, 
e^{2\pi i \frac{b}{N}(v+a)}\, X^{(N)}(v,a;\tau,z).
\label{Xva def 2}
\end{equation}

The main formula that we would like to prove in this section is 
\begin{equation}
\mathboxit{
X^{(N)}(v,a;\tau,z) = \hchid^{(N,1)}(v,a;\tau,z) + \hchid^{(N,1)}(N-v, a+v;\tau,z).
}
\label{X hchid}
\end{equation}



In order to achieve this formula 
we first consider the elliptic genera of $SL(2)_{N+2}/U(1)$-supercoset with $ \hc = 1 + \frac{2}{N} $
 \cite{Troost,ES-NH}.
We set 
\begin{equation}
\cZ(\tau,z) := \hcK^{(2N)}\left(\tau, \frac{z}{N}\right)
\frac{\th_1(\tau,z)}{i\eta(\tau)^3},
\label{ell genus 1}
\end{equation}
and 
\begin{eqnarray}
\tcZ(\tau,z) &:=& \frac{1}{N} \sum_{a,b\in \bz_N}\, s_{(a,b)} \cdot \cZ (\tau,z)
\nn
&\equiv & \frac{1}{N} \sum_{a,b\in \bz_N}\,q^{\frac{a^2}{N}} 
y^{\frac{2a}{N}} e^{2\pi i \frac{ab}{N}} \,
\hcK^{(2N)}\left(\tau, \frac{z+a\tau+b}{N}\right)
\frac{\th_1(\tau,z)}{i\eta(\tau)^3}.
\label{ell genus 2}
\end{eqnarray}
Here, 
$\tcZ(\tau,z)$ is identified with the elliptic genus of axial supercoset of $SL(2)_{N+2}/U(1)$ (`cigar' \cite{2DBH}), 
while $\cZ(\tau,z)$ is associated with the vector supercoset;
\begin{eqnarray*}
[\mbox{vector} ~ SL(2)/U(1)] &\cong & [\mbox{$\bz_N$-orbifold of axial} ~ SL(2)/U(1)] \\
&\cong & [\mbox{$N$-fold cover of `trumpet'}],
\end{eqnarray*}
as is shown in \cite{orb-ncpart}.
The following identities play crucial roles; 
%
\begin{eqnarray}
&& \cZ_{(a,b)}(\tau, z) :=  s_{(a,b)}\cdot \cZ (\tau,z)  
= \sum_{v=0}^{N-1} \, e^{2\pi i \frac{b}{N}(v+a)} \hchid^{(N,1)}(v,a;\tau,z),
\label{Zab hchid}
\\
&& \tcZ_{(a,b)}(\tau, z) :=  s_{(a,b)}\cdot \tcZ (\tau,z)
\left( \, \equiv
\frac{1}{N}\sum_{\al,\beta\in \bz_N}\, e^{-2\pi i \frac{1}{N}(a \beta- b \al)}\,
\cZ_{(\al,\beta)}(\tau,z)
\,
\right) 
\nn
&&
\hspace{2cm}
= \sum_{v=0}^{N-1} \, e^{2\pi i \frac{b}{N}(v+a)} \hchid^{(N,1)}(N- v,a+v;\tau,z),
\label{Z'ab hchid}
\end{eqnarray}
which are  proven in \cite{ES-NH,orb-ncpart}.

We also note that the IR-parts of $\cZ_{(a,b)}$ and $\tcZ_{(a,b)}$
are given as 
\begin{eqnarray}
[ \cZ_{(a,b)} (\tau,z) ] & =& \delta^{(N)}_{a,0} \, 
\left\{
\sum_{v=1}^{N-1} 
\, e^{2\pi i \frac{b}{N}v} y^{-\frac{1}{2}+\frac{v}{N}}
+ \frac{1}{2} \left(y^{-\frac{1}{2}} + y^{\frac{1}{2}}\right)
\right\}
\label{IR Zab}
\\
\left\lb \tcZ_{(a,b)}
(\tau, z) \right\rb
&=& y^{-\frac{1}{2}+\frac{[a]}{N}} 
+ \delta_{a,0}^{(N)}\,
\frac{1}{2} \left(
y^{\frac{1}{2}} - y^{-\frac{1}{2}}
\right).
\label{IR tZab}
\end{eqnarray}


With these preparations we shall  prove the next identity;
\begin{equation}
\mathboxit{
\cZ(\tau,z) + \tcZ(\tau,z) = \Phi^{(N)}(\tau,z),}
\label{Z tZ Phi}
\end{equation}
from which the identity \eqn{X hchid} is readily derived by using \eqn{Zab hchid}, \eqn{Z'ab hchid} as well as the definition $X^{(N)}(v,a)$ \eqn{Xva def}.

~


\noindent
{\bf [proof of \eqn{Z tZ Phi}]}

We set $\Phi^{(N) '}(\tau,z):= \cZ(\tau,z) + \tcZ(\tau,z) $, 
and will prove $\Phi^{(N) '}(\tau,z) = \Phi^{(N)}(\tau,z)$.

We first enumerate relevant properties of $\Phi^{(N) '}(\tau,z)$;


~

\begin{description}

\item[(i) modular and spectral flow properties :]

~


We first note that $\Phi^{(N) '}(\tau,z)$ possesses the 
expected modular and spectral flow properties;
\begin{eqnarray}
&& \Phi^{(N) '}(\tau+1,z) = \Phi^{(N) '}(\tau,z), 
\hspace{1cm} 
\Phi^{(N) '}\left(-\frac{1}{\tau},\frac{z}{\tau} \right)
\label{modular Phi'} = e^{i\pi \frac{\hc}{\tau}z^2}\,
\Phi^{(N) '}(\tau,z),
\\
&& s_{(Na,Nb)}\cdot \Phi^{(N) '} (\tau,z) = \Phi^{(N) '} (\tau,z), \hspace{1cm}
(\any a,b \in \bz).
\end{eqnarray}
They are shown from the same properties of $\cZ(\tau,z)$ as well as the fact 
that $s_{(a,b)}$ acts modular covariantly.

~


\item[(ii) holomorphicity :]

~


Recall the fact  that $\cZ(\tau,z)$ is written as  
\begin{equation}
\cZ(\tau,z) = \frac{\th_1(\tau,z)}{i\eta(\tau)^3} \,
\left[
\cK^{(2N)} \left(\tau,\frac{z}{N}\right)
-\frac{1}{2} \sum_{m\in \bz_N}\, R_{m,N}(\tau) 
\Th{m}{N}\left(\tau,\frac{2z}{N}\right)
\right],
\end{equation}
and the second term is non-holomorphic. 
Let us consider the `$\bz_N$-orbifold action', that is 
$$
\frac{1}{N}\sum_{a,b\in \bz_N} s_{(a,b)} \cdot \lb \, \ast \, \rb,
$$
 to this non-holomorphic correction term.
Because of the simple identity 
\begin{equation}
\frac{1}{N} \sum_{a,b\in \bz_N}\, 
q^{\frac{a^2}{N}} y^{\frac{2a}{N}}
e^{2\pi i \frac{ab}{N}}\, \Th{m}{N}
\left(\tau,\frac{2}{N}(z+a\tau+b) \right) 
= \Th{-m}{N} \left(\tau, \frac{2z}{N}\right), 
\label{orbifolding theta}
\end{equation}
we obtain 
\begin{eqnarray*}
[\mbox{non-hol. corr.  term}] 
& \stackrel{\msc{$\bz_N$-orbifolding}}{\Longrightarrow} &
-\frac{1}{2} \sum_{m\in \bz_N}\, R_{m,N}(\tau) 
\Th{-m}{N}\left(\tau,\frac{2z}{N}\right)\, \frac{\th_1(\tau,z)}{i\eta(\tau)^3}
\\
&=& -\frac{1}{2} \sum_{m\in \bz_N}\, R_{-m,N}(\tau) 
\Th{m}{N}\left(\tau,\frac{2z}{N}\right)\, \frac{\th_1(\tau,z)}{i\eta(\tau)^3}
\\
&=& \frac{1}{2} \sum_{m\in \bz_N}\, \left\{ R_{m,N}(\tau) - 2\delta^{(N)}_{m,0}
\right\}
\Th{m}{N}\left(\tau,\frac{2z}{N}\right)\, \frac{\th_1(\tau,z)}{i\eta(\tau)^3}
\\
&&
\hspace{2.5cm} (\because ~ R_{-m,N}(\tau) = -R_{m,N}(\tau) + 2 \delta^{(N)}_{m,0} ).
\end{eqnarray*}
Therefore, potential non-holomorphic terms in $\Phi^{(N)'}(\tau,z)$
are strictly canceled out, and we can rewrite it as 
\begin{eqnarray}
\Phi^{(N) '}(\tau,z) &=& \frac{\th_1(\tau,z)}{i\eta(\tau)^3}
\left[
\cK^{(2N)}\left(\tau,\frac{z}{N} \right) 
+ \frac{1}{N} \sum_{a,b\in \bz_N} \, q^{\frac{a^2}{N}} y^{\frac{2a}{N}}
e^{2\pi i \frac{ab}{N}}\, \cK^{(2N)}\left(\tau,\frac{z+a\tau+b}{N} \right)
\right.
\nn
&&
\hspace{7.5cm}
\left.
- \Th{0}{N} \left(\tau, \frac{2z}{N} \right) 
\right].
\label{Phi' 2}
\end{eqnarray}
This is manifestly holomorphic.

~


\item[(iii) IR-part :]

~


Recall
\begin{equation}
[\cZ(\tau,z)] = \frac{1}{2} \left(y^{\frac{1}{2}} 
+ y^{-\frac{1}{2}} \right) + 
\sum_{v=1}^{N-1} y^{-\frac{1}{2} + \frac{v}{N}},
\hspace{1cm}
[\tcZ(\tau,z)] = \frac{1}{2} \left(y^{\frac{1}{2}} 
+ y^{-\frac{1}{2}} \right),
\end{equation}
and thus
\begin{equation}
\left[\Phi^{(N)'}(\tau,z)\right] = \sum_{v=0}^N \,
y^{-\frac{1}{2} + \frac{v}{N}}
 \left(\, \equiv  \left[\Phi^{(N)}(\tau,z)\right] \, \right) .
\label{IR Phi'}
\end{equation}
Moreover, we can show 
\begin{equation}
\left[
\Phi^{(N)\,'}_{(a,b)}(\tau,z) 
\right]
= \delta_{a,0}^{(N)} \,  \sum_{v=1}^{N} \, e^{2\pi i \frac{b}{N}v}
y^{-\frac{1}{2} + \frac{v}{N}}
+   y^{- \frac{1}{2} +\frac{[a]}{N}}
\left( \equiv  \left[ \Phi^{(N)}_{(a,b)}(\tau,z) \right] \, \right),
\label{IR Phi' ab}
\end{equation}
due to \eqn{IR Zab}, \eqn{IR tZab} and \eqn{IR Phi ab}.

\end{description}

~


In this way, we can conclude that 
both of $\Phi^{(N)}(\tau,N z)$ and  $\Phi^{(N)\,'}(\tau,N z)$
are holomorphic Jacobi forms of weight 0 and index 
$\frac{ N^2 \hc}{2} \equiv \frac{N(N+2)}{2}$ which 
share the IR-part,
namely, 
\begin{eqnarray*}
&& \Phi^{(N)}(\tau,N z), \, \Phi^{(N)\, '}(\tau,N z) \in 
\cJ \left[ 0, \frac{N(N+2)}{2}\right],
\\
&& \left\lb \Phi^{(N)}(\tau,N z) \right\rb
= \left\lb \Phi^{(N)\, '}(\tau,N z) \right\rb = \sum_{v=0}^N 
y^{-\frac{N}{2}+v}.
\end{eqnarray*}

Consequently, if setting 
\begin{equation}
F(\tau,z) := \frac{\Phi^{(N)\, '}(\tau, Nz)}{ \Phi^{(N)}(\tau,Nz)}, 
\label{def F}
\end{equation}
then $F(\tau,z)$ is an elliptic, modular invariant function with the IR-part
$[F(\tau,z)] =1$. 
Thus, if we would succeed in proving the holomorphicity of $F(\tau,z)$
for $\any \tau \in \bh \cup \{i\infty\}$, $\any z \in \bc $, 
we can conclude $F(\tau,z) \equiv 1$ and the proof will be completed.

$\Phi^{(N)\, '}(\tau,Nz)$ is obviously holomorphic for $\any \tau \in \bh \cup \{i\infty\}$, $\any z \in \bc $, 
and $\Phi^{(N)}(\tau,Nz)$ has simple zeros only at $N(N+2)$ points 
\begin{equation}
z_{a,b} := \frac{a\tau+b}{N+1} , ~~~ a,b =0,1,\ldots, N, ~~ (a,b) \neq (0,0) ,
\label{z ab}
\end{equation}
in the fundamental region of double periodicity of $F$. 
Hence the function $F(\tau,z)$ at most possesses simple poles at $z=z_{a,b}$ in the fundamental region, and  
the following lemma is enough for completing the proof;

~


\noindent
{\bf [Lemma]} ~~ 
All the residues of $F(\tau,z)$ at $z=z_{a,b}$ vanish.


~

\noindent
$\because~)$ ~ Let us denote 
\begin{equation}
R_{a,b}(\tau) := \mbox{Res}_{z=z_{a,b}} \left\lb F(\tau,z)\right\rb
\equiv \frac{1}{2\pi i} \oint_{C_{a,b}(\tau)} dz\, F(\tau,z),
\label{Rab}
\end{equation}
where $C_{a,b}(\tau)$ is a small contour encircling $z_{a,b}$. 
Because of the modular invariance of $F(\tau,z)$, 
we find the modular properties of $R_{a,b}$ as 
\begin{equation}
R_{a,b}(\tau+1) = R_{a,a+b}(\tau), \hspace{1cm}
R_{a,b}\left(-\frac{1}{\tau} \right) = \frac{1}{\tau} \, R_{b,-a}(\tau).
\end{equation}
In fact, the first formula is trivial, and the second formula is proven as follows;
\begin{eqnarray*}
R_{a,b}\left(-\frac{1}{\tau} \right) &=& \frac{1}{2\pi i } \oint_{C_{a,b}\left(-\frac{1}{\tau} \right)}dz \,
F\left(-\frac{1}{\tau}, z \right) 
\\
&=& \frac{1}{2\pi i } \oint_{C_{b,-a}\left(\tau \right)}\frac{dz'}{\tau} \,
F\left(-\frac{1}{\tau}, \frac{z'}{\tau} \right) \hspace{1cm}
(z':= \tau z)
\\
&=& 
\frac{1}{\tau} \frac{1}{2\pi i } \oint_{C_{b,-a}\left(\tau \right)}dz'\,
F\left(\tau, z' \right) \hspace{1cm} (\because ~ \mbox{modular invariance of $F$})
\\
&=& \frac{1}{\tau} R_{b,-a} (\tau).
\end{eqnarray*}
It is obvious that $R_{a,b}(\tau)$ is holomorphic over $\bh$, since $z=z_{a,b}$ is at most a simple pole and 
$\Phi^{(N) '}(\tau, Nz)$ is holomorphic for $\any \tau \in \bh$, $\any z \in \bc$.
Moreover, one can show 
\begin{equation}
\lim_{\tau\rightarrow i\infty}\, \left| R_{a,b}(\tau) \right| < \infty, ~~~  (\any a,b),
\label{Rab cusp}
\end{equation}
with the helps of \eqn{IR Phi'} and  \eqn{IR Phi' ab}.
The proof of \eqn{Rab cusp} is straightforward, and we shall present it in Appendix C.


Now, we define the `modular orbit' $\cO_r \subset \bz_{N+1} \times \bz_{N+1}$
for 
$\any r \in D(N+1) \equiv \{r=1,2, \ldots, N~;~ r\, \left| \, (N+1) \right. \}$
as 
\begin{equation}
\cO_r := \left\{ (a,b) \in \bz_{N+1} \times \bz_{N+1} ~:~ 
(a,b) \equiv (r,0) A ~ (\mbox{mod}~ (N+1) \times (N+1)),~~ {}^{\exists}
A \in SL(2,\bz)  \right\}.
\nonumber
\end{equation}
Then, 
$$
\bz_{N+1}\times \bz_{N+1} -  \{(0,0)\} = \coprod_{r \in D(N+1)}\, \cO_r,
$$
holds, and 
\begin{equation}
R^{(r,k)}(\tau) := \sum_{(a,b) \in \cO_r} \, \left(R_{a,b}(\tau)\right)^{2k},
\label{R rk}
\end{equation}
should be a modular form of weight $-2k$.
Therefore, 
$R^{(r,k)}(\tau)$ has to vanish everywhere over $ \bh \cup \{i\infty\}$
for arbitrary $r \in D(N+1)$ 
and $k\in \bz_{>0}$. 
This is sufficient to conclude that 
$R_{a,b}(\tau) \equiv 0$ for every pole $z_{a,b}$.

~

In this way, the identity \eqn{Z tZ Phi} has been proven, leading to 
the main formula \eqn{X hchid}.
~~ {\bf (Q.E.D)}

~


\section{Physical Interpretation}

In this section we try to make a physical interpretation of the main result given above, that is, 
the identities \eqn{X hchid} or \eqn{Z tZ Phi}. 
In other words, we discuss what superconformal system leads to $\Phi^{(N)}(\tau,z)$ \eqn{Phi N} 
as its elliptic genus.

~

\subsection{Some Properties of the Function $X^{(N)}(v,a)$}

We first note several helpful facts about the expected `building block' $X^{(N)}(v,a)$ \eqn{Xva def}:
\begin{description}
\item[1. modular and spectral flow properties :]

~

Once we achieve the identity \eqn{X hchid}, we can readily derive the formulas of modular and
spectral flow transformations of $X^{(N)}(v,a)$  
by using  \eqn{S hchid K=1}, \eqn{T hchid K=1} and \eqn{sflow hchid K=1},
which are written  as 
\begin{eqnarray}
\hspace{-1cm}
X^{(N)} \left(v,a ; - \frac{1}{\tau}, \frac{z}{\tau}\right)
&=& e^{i\pi \frac{\hc}{\tau}z^2}\, \sum_{v'=0}^{N-1} \,\sum_{a'\in \bz_N}\,
\frac{1}{N} e^{2\pi i \frac{vv' - (v+2a)(v'+2a')}{2N}}
\, X^{(N)} (v',a';\tau,z)
\nn
&\equiv & 
e^{i\pi \frac{\hc}{\tau}z^2}\, \sum_{v'=0}^{N-1} \,\sum_{a'\in \bz_N}\,
\frac{1}{N} e^{- 2\pi i \frac{(v+2a)(v'+2a')}{2N}} \cos \left(\frac{\pi v v'}{N}\right)
\, X^{(N)} (v',a';\tau,z)
\nn
&&
\label{S X}
\\
\hspace{-1cm}
X^{(N)} \left(v,a ; \tau+1, z \right)
&=& e^{2\pi i \frac{a}{N} \left(v+ a \right)}\,
X^{(N)}\left(v,a ; \tau, z \right).
\label{T X}
\\
\hspace{-1cm}
X^{(N)} (v,a;\tau,z+r\tau+s) 
&=& (-1)^{r+s} e^{2\pi i \frac{v+2a}{N}s}
q^{-\frac{\hc}{2}r^2} y^{-\hc r}\, X^{(N)} (v,a+r;\tau,z),
~~~ (\any r,s \in \bz),
\label{sflow X}
\end{eqnarray}
In other words, 
\begin{equation}
s_{(r,s)}\cdot X^{(N)} (v, a) = e^{2\pi i \frac{v+2a +r}{N} s}\, X^{(N)} (v, a +r), ~~~ (\any r,s \in \bz).
\label{sflow X 2}
\end{equation}

As a consistency check, one may also derive these formulas directly from the `Fourier expansion relation'
\eqn{Xva def 2}.

~


\item[2. manifestly  holomorphic expression : ]

~
 
In place of \eqn{X hchid},  $X^{(N)}(v,a)$ can be rewritten in a manifestly holomorphic expression;
\begin{equation}
X^{(N)}(v,a;\tau,z) = \chid^{(N,1)}(v,a;\tau,z) + \chid^{(N,1)}(N-v,a+v; \tau,z) .
\label{X chid}
\end{equation} 
In fact, the non-holomorphic correction terms in the R.H.S of \eqn{X hchid} are found to be canceled out precisely, as in 
\eqn{Phi' 2}. 
One can achieve this identity by making the Fourier expansion of \eqn{Phi' 2} and recalling the identity $\Phi^{(N)}(\tau,z) = \Phi^{(N)\, '}(\tau,z)$
proven in the previous section.

~


\item[3. interpretation as `analytic continuation' of the character of $\cN=2$ minimal model ]

~

It would be worthwhile to recall basic facts of the $\cN=2$ minimal model, namely, 
the $SU(2)_{N-2}/U(1)$-supercoset which has $\hc = 1-\frac{2}{N}$. 
As we mentioned at the beginning of this paper, 
the elliptic genus of the minimal model is given as \cite{Witten-EG minimal}
\begin{equation}
\cZ^{(N-2)}_{\msc{min}} (\tau,z) = \frac{\th_1\left(\tau, \frac{N-1}{N} z\right)}{\th_1\left(\tau, \frac{z}{N}\right)}
= \sum_{\ell=0}^{N-2} \, \ch{(\stR)}{\ell, \ell+1} (\tau,z),
\label{minimal ell genus}
\end{equation}
where $\ch{(\stR)}{\ell, m} (\tau,z)$ denotes the $\tR$-character of the $\cN=2$ minimal model which has 
the Witten index 
\begin{equation}
\ch{(\stR)}{\ell, m} (\tau,0) = \delta^{(2N)}_{m, \ell+1} - \delta^{(2N)}_{m, -(\ell+1)}. 
\end{equation}
Moreover, the spectrally flowed elliptic genus is Fourier expanded in terms of the $\tR$-characters as follows
($a,b\in \bz_N$);
\begin{eqnarray}
\cZ^{(N-2)}_{\msc{min}\, (a,b)} (\tau,z) &\equiv & (-1)^{a+b} q^{\frac{\hc}{2} a^2} y^{\hc a} e^{-2\pi i \frac{ab}{N}}\,
\cZ^{(N-2)}_{\msc{min}}(\tau, z+a\tau+b) 
\nn
&=& \sum_{\ell=0}^{N-2} \, e^{2\pi i \frac{b}{N} (\ell+1 -a)}\, \ch{(\stR)}{\ell, \ell+1 -2a} (\tau,z).
\label{exp minimal ell genus}
\end{eqnarray}
In this way, by comparing \eqn{Xva def 2} with \eqn{exp minimal ell genus}
one would find a similarity between the function $X^{(N)}(v,a;\tau,z)$ and the minimal character 
$\ch{(\stR)}{\ell, m} (\tau,z)$ with the correspondence
\begin{equation}
N~\longrightarrow~ -N , \hspace{1cm}
\ell  ~\longrightarrow ~ -v-1 , \hspace{1cm}  m  ~\longrightarrow ~ - v- 2a.
\label{correspondence to minimal}
\end{equation}

Furthermore, let us recall the modular transformation formula of $\ch{(\stR)}{\ell, m} (\tau,z)$;
\begin{eqnarray}
\hspace{-5mm}
\ch{(\stR)}{\ell, m} (\tau +1,z) &=& e^{2\pi i \left\{  \frac{\ell(\ell+2)-m^2}{4N} + \frac{1}{8} -\frac{N-2}{8N} \right\} }\, 
\ch{(\stR)}{\ell, m} (\tau ,z) 
\nn
&=&  e^{2\pi i \frac{(\ell+1)^2- m^2}{4N}}\, \ch{(\stR)}{\ell, m} (\tau ,z) , \hspace{1cm} (\mbox{with}~ m\equiv \ell+1 +2a) ,
\label{T minimal}
\\
\hspace{-5mm} 
\ch{(\stR)}{\ell,m} \left(-\frac{1}{\tau}, \frac{z}{\tau} \right) &=& (-i) e^{i\pi \frac{\hc}{\tau} z^2}\, 
\sum_{\ell'=0}^{N-2}\, \sum_{\stackrel{m'\in \bz_{2N}}{\ell'+m' \in 2\bz +1}} \, 
\nn
&& 
\hspace{5mm}
\times
\sqrt{\frac{2}{N}} \sin \left(\frac{\pi (\ell+1)(\ell'+1)}{N}\right)\,
\frac{1}{\sqrt{2N}} e^{2\pi i \frac{m m'}{2N}} \, \ch{(\stR)}{\ell', m'} (\tau ,z)
\nn
&=&  e^{i\pi \frac{\hc}{\tau} z^2}\, 
\sum_{\ell'=0}^{N-2}\,  \sum_{\stackrel{m'\in \bz_{2N}}{\ell'+m' \in 2\bz +1}} \, 
\frac{1}{N} e^{-2\pi i \frac{(\ell+1)(\ell'+1)-m m'}{2N}} \, \ch{(\stR)}{\ell', m'} (\tau ,z).
\label{S minimal}
\end{eqnarray}
The first line of \eqn{S minimal} is familiar formula and we have made use of the property of minimal character;
\begin{equation}
\ch{(\stR)}{N-2-\ell, m+N} (\tau,z) = - \ch{(\stR)}{\ell, m} (\tau,z),
\label{fi minimal}
\end{equation}
to derive the second line\footnote
{
We have an obvious identity for $X^{(N)}(v,a;\tau,z)$
$$
X^{(N)}(N-v,v+a;\tau,z) = X^{(N)}(v,a;\tau,z),
$$
which again corresponds to \eqn{fi minimal} under \eqn{correspondence to minimal}
up to the overall sign.
}.
The modular transformation formulas \eqn{T X}, \eqn{S X}   nicely correspond to \eqn{T minimal}, \eqn{S minimal}  
under the formal replacement \eqn{correspondence to minimal}. 
In this way, one would regard $X^{(N)}(v,a;\tau,z)$ as a formal `analytic continuation' of 
the minimal character $\ch{(\stR)}{\ell, m} (\tau,z)$.


\end{description}

~


\subsection{`Compactified $SL(2)/U(1)$-Supercoset' in the $(\tR,\tR)$-sector}

Now, let us discuss what is the superconformal model whose modular invariant 
is build up from the functions $X^{(N)}(v,a)$ \eqn{X hchid}.

Let $\cH^{(\sR)}_A$ be the Hilbert space of the axial supercoset of $SL(2)_{N+2}/U(1)$ in the $(\R,\R)$-sector, while 
$\cH^{(\sR)}_V$ be that of the vector supercoset. 
In other words, $\cH^{(\sR)}_A$ corresponds to the cigar-type superconformal model with the asymptotic radius $\sqrt{N\al'}$, whereas 
$\cH^{(\sR)}_V$ should be associated with 
$$
[\mbox{cigar}]/\bz_N \cong [\mbox{$N$-fold cover of trumpet}],
$$
as is already mentioned in section 2.
This means that the torus partition functions in the $(\tR, \tR)$-sector
 of these theories are schematically written as ($\hc \equiv 1 + \frac{2}{N}$, $\tau\equiv \tau_1+i\tau_2$, $z\equiv z_1 + iz_2$, and 
$F_L$, $F_R$ denotes the fermion number operators $\mod \, 2$)
\begin{eqnarray}
Z^{(\stR)}_V(\tau,\bar{\tau};z,\bar{z}) &\equiv & e^{-2\pi \hc \frac{z_2^2}{\tau_2}}\,
\tr_{\cH^{(\sR)}_V} \left[\, (-1)^{F_L+F_R}\, q^{L_0-\frac{\hc}{8}} \overline{q^{\tL_0-\frac{\hc}{8}}} \, e^{2\pi i z J_0} \overline{e^{2\pi i z \tJ_0}} \right]
\nn
&=&  e^{-2\pi \hc \frac{z_2^2}{\tau_2}}\, 
\sum_{v=0}^{N-1} \sum_{a\in \bz_N}\, \hchid^{(N,1)}(v,a;\tau,z) \,\left[\hchid^{(N,1)}(v,a;\tau,z)\right]^* + \left[\mbox{cont. terms}\right],
\label{part fn vector}
\\
Z^{(\stR)}_A(\tau,\bar{\tau};z,\bar{z}) &\equiv &  e^{2\pi \hc \frac{z_1^2}{\tau_2}}\,
\tr_{\cH_A^{(\sR)}} \left[\, (-1)^{F_L+F_R}\, q^{L_0-\frac{\hc}{8}} \overline{q^{\tL_0-\frac{\hc}{8}}} \, e^{2\pi i z J_0} \overline{e^{2\pi i z \tJ_0}} \right]
\nn
&=& e^{2\pi \hc \frac{z_1^2}{\tau_2}}\,\sum_{v=0}^{N-1} \sum_{\stackrel{a_L, a_R \in \bz_N}{v+ a_L + a_R \in N\bz}}\, 
\hchid^{(N,1)}(v,a_L;\tau,z) \,\left[\hchid^{(N,1)}(v,a_R;\tau,z)\right]^* + \left[\mbox{cont. terms}\right]
\nn
&=& e^{2\pi \hc \frac{z_1^2}{\tau_2}}\,\sum_{v=0}^{N-1} \sum_{a\in \bz_N}\, 
\hchid^{(N,1)}(v,-v-a;\tau,z) \,\left[\hchid^{(N,1)}(v,a;\tau,z)\right]^* + \left[\mbox{cont. terms}\right].
\nn
&&
\label{part fn axial}
\end{eqnarray}
The discrete part of \eqn{part fn vector} obviously looks 
the diagonal modular invariant, whereas that of \eqn{part fn axial} can be regarded as the anti-diagonal one
with respect to the `minimal model like' quantum number $m\equiv v + 2a$ adopted in \cite{orb-ncpart}\footnote
    {The quantum number $m$ labels the $U(1)_R$-charges appearing in the spectral flow orbit defining $\chid^{(N,1)}(v,a)$.
As mentioned in \cite{orb-ncpart}, the axial type \eqn{part fn axial} is anti-diagonal 
only in the case of integer levels ({\em i.e.} $K=1$), while the vector 
type \eqn{part fn vector} is always diagonal. } .

The `anomaly factors' $e^{-2\pi \hc \frac{z_2^2}{\tau_2}}$, $e^{2\pi \hc \frac{z_1^2}{\tau_2}}$, which ensures the modular invariance,  
originate precisely from the path-integrations, and they differ due to the gauged WZW actions of vector and axial types \cite{ES-NH,orb-ncpart}.
Note that these factors just get the common form $e^{2\pi \hc \frac{z_1^2}{\tau_2}}$, if we replace $z$ with $-z$ in the axial model. 
Thus it would be useful to rewrite \eqn{part fn axial} as 
\begin{eqnarray}
Z^{(\stR)}_A(\tau,\bar{\tau};-z,\bar{z})
&=&  e^{-2\pi \hc \frac{z_2^2}{\tau_2}}\,\sum_{v=0}^{N-1} \sum_{a\in \bz_N}\, 
\hchid^{(N,1)}(v,-v-a;\tau, - z) \,\left[\hchid^{(N,1)}(v,a;\tau,z)\right]^* + \left[\mbox{cont. terms}\right]
\nn
&=& e^{-2\pi \hc \frac{z_2^2}{\tau_2}}\,\sum_{v=0}^{N-1} \sum_{a\in \bz_N}\, 
\hchid^{(N,1)}(N-v,v+a;\tau,z) \,\left[\hchid^{(N,1)}(v,a;\tau,z)\right]^* + \left[\mbox{cont. terms}\right],
\nn
&&
\label{part fn axial 2}
\end{eqnarray} 
where we made use of the identity
\begin{equation}
\hchid^{(N,1)}(v,a;\tau, -z) = \hchid^{(N,1)}(N-v,-a;\tau,z),
\label{cc rel hchid}
\end{equation}
in the second line.

Moreover, it is found that the continuous terms appearing in \eqn{part fn vector} and \eqn{part fn axial 2} 
are written in the precisely same functional form {\em with inverse sign}. 
We will prove  this  fact in the next subsection, 
and here address our main result in this section;
\begin{equation}
\mathboxit{
\begin{array}{lll}
Z^{(\stR)}_{cSL(2)/U(1)}(\tau,\bar{\tau};z,\bar{z}) &:=& \dsp Z^{(\stR)}_V(\tau,\bar{\tau};z,\bar{z}) + Z^{(\stR)}_A(\tau,\bar{\tau};-z,\bar{z}) 
\\
&=& \dsp \frac{1}{2} e^{-2\pi \hc \frac{z_2^2}{\tau_2}} \, \sum_{v=0}^{N-1} \sum_{a\in \bz_N}\,  
\left|X^{(N)}(v,a;\tau,z)\right|^2.
\end{array}
}
\label{part cSL(2)/U(1)}
\end{equation}
Note that, even though  both $Z^{(\stR)}_V$ and $Z^{(\stR)}_A$ include 
contributions of continuous characters with 
non-trivial coefficients showing IR-divergence, the combined partition function 
\eqn{part cSL(2)/U(1)} is written in terms only of a finite number of holomorphic building blocks 
$X^{(N)}(v,a;\tau,z)$ given in \eqn{X hchid} or \eqn{X chid}.
This fact strongly suggests that the modular invariant partition function \eqn{part cSL(2)/U(1)} would define 
an $\cN=2$ superconformal model with $ \hc= 1+ \frac{2}{N}$ 
associated with some {\em compact} background. 
Therefore, we tentatively call this model as the  `compactified $SL(2)/U(1)$-supercoset model' here. 
It is obvious that the elliptic genus of this model is given by the holomorphic function $\Phi^{(N)}(\tau,z)$
\eqn{Phi N};
\begin{equation}
\cZ_{cSL(2)/U(1)}(\tau, z) = \Phi^{(N)}(\tau,z) \equiv \frac{\th_1\left(\tau,\frac{N+1}{N} z\right)}
{\th_1\left(\tau,\frac{1}{N}z\right)}.
\label{EG cSL(2)/U(1)}
\end{equation}

~


\subsection{Cancellation of Continuous Terms}

In this subsection, we prove the precise cancellation of  continuous terms 
potentially appearing in the first line of \eqn{part cSL(2)/U(1)}

We start with the explicit form of the partition function $Z^{(\stR)}_V$, 
which is evaluated in \cite{orb-ncpart} by means of the path-integration ($u\equiv s_1 \tau+ s_2$);
\begin{eqnarray}
\hspace{-5mm}
Z^{(\stR)}_{V}(\tau,\bar{\tau} ;z, \bar{z};\ep) &=& 
N\, e^{-\frac{2\pi}{\tau_2} \frac{N+4}{N} z_2^2}\,
\sum_{n_1,n_2\in \bz}\, \int_{\Sigma(z,\ep)} \frac{d^2 u}{\tau_2} \, 
\left|
\frac{\th_1\left(\tau,u+\frac{N+2}{N}z\right)}{\th_1\left(\tau, u+\frac{2}{N}z\right)}
\right|^2 \,
\nn
&& \hspace{3cm} \times 
e^{-4\pi \frac{u_2 z_2}{\tau_2}} 
e^{-\frac{\pi N}{\tau_2} \left|n_1\tau+n_2\right|^2} 
 e^{2\pi i N (n_1 s_2- n_2 s_1)}.
\label{part fn V}
\end{eqnarray}
Here we introduced the IR-regularization adopted in \cite{ES-NH,orb-ncpart},
which removes the singularity of integrand originating from the non-compactness of target space. 
Namely we set 
\begin{equation}
\Sigma(z, \ep) \equiv \Sigma \,  \setminus \, 
\left\{u= s_1 \tau + s_2 ~~;~~
-\frac{\ep}{2} - \frac{2}{k} \zeta_1 < s_1 < 
\frac{\ep}{2} - \frac{2}{k} \zeta_1,   ~ ~ 0< s_2 < 1  \right\},
\label{Sigma ep}
\end{equation}
where we set $z\equiv \zeta_1 \tau + \zeta_2$, 
$\zeta_1, \zeta_2 \in \br$, and $\ep(>0)$ denotes 
the regularization parameter.

In the same way, the axial partition function $Z^{(\stR)}_A$, which describes 
the cigar background, is written as \cite{ES-NH}
\begin{eqnarray}
Z^{(\stR)}_{A}(\tau,\bar{\tau} ; z, \bar{z}; \ep) &=& 
N e^{\frac{2\pi}{\tau_2} \left( \hc \left|z\right|^2 - \frac{N+4}{N} z_2^2\right)}\,
\sum_{m_1,m_2\in \bz}\, \int_{\Sigma(z,\ep)} \frac{d^2 u}{\tau_2} \, 
\left|
\frac{\th_1\left(\tau,u+\frac{N+2}{N}z\right)}{\th_1\left(\tau, u+\frac{2}{N}z\right)}
\right|^2 \,
\nn
&& 
\hspace{3cm} 
\times
e^{-4\pi \frac{u_2z_2}{\tau_2}} \, e^{-\frac{\pi N }{\tau_2} \left|m_1\tau+m_2+ u\right|^2}.
\label{part fn A}
\end{eqnarray}

As demonstrated in \cite{ES-NH},  one can make the `character decompositions' of partition functions
\eqn{part fn V}, \eqn{part fn A}.
They are schematically expressed as 
\begin{eqnarray}
&& Z^{(\stR)}_V(\tau,\bar{\tau}; z, \bar{z} ; \ep) = 
 e^{-2\pi \hc \frac{z_2^2}{\tau_2}}\, 
\left[ \sum_{v=0}^{N-1} \sum_{a\in \bz_N}\, \hchid^{(N,1)}(v,a;\tau,z) \,\left[\hchid^{(N,1)}(v,a;\tau,z)\right]^* 
\right.
\nn
&& 
\hspace{1cm}
+ \sum_{m\in \bz_{2N}}\, \int dp_L\, \int dp_R\, 
\left\{\rho_1(p_L,p_R, m ; \ep ) 
\chic^{(N,1)}(p_L,m;\tau,z) \left[\chic^{(N,1)}(p_R ,m;\tau,z)\right]^* 
\right.
\nn
&& 
\hspace{3cm}
\left. \left. 
+ \rho_2(p_L,p_R, m ; \ep ) \chic^{(N,1)}(p_L,m;\tau,z) 
\left[\chic^{(N,1)}(p_R ,m+N ;\tau,z)\right]^* \right\}
\right],
\label{ch exp V}
\end{eqnarray}
\begin{eqnarray}
&& Z^{(\stR)}_A(\tau,\bar{\tau}; z, \bar{z} ; \ep) = 
 e^{2\pi \hc \frac{z_1^2}{\tau_2}}\, 
\left[ \sum_{v=0}^{N-1} \sum_{a\in \bz_N}\, \hchid^{(N,1)}(v,-v-a;\tau,z) \,\left[\hchid^{(N,1)}(v,a;\tau,z)\right]^* 
\right.
\nn
&& 
\hspace{1cm}
+ \sum_{m\in \bz_{2N}}\, \int dp_L\, \int dp_R\, 
\left\{\rho'_1(p_L,p_R, m ; \ep ) \chic^{(N,1)}(p_L, -m;\tau,z) 
\left[\chic^{(N,1)}(p_R ,m;\tau,z)\right]^* 
\right.
\nn
&& 
\hspace{3cm}
\left. \left. 
+ \rho'_2(p_L,p_R, m ; \ep ) \chic^{(N,1)}(p_L,- m;\tau,z) 
\left[\chic^{(N,1)}(p_R ,m+N ;\tau,z)\right]^* \right\}
\right],
\label{ch exp A}
\end{eqnarray}
where 
$\chic^{(N,1)}(p,m)$
denotes the extended continuous character \eqn{chic} explicitly written as 
$$
\chic^{(N,1)}(p,m;\tau,z) = q^{\frac{p^2}{4N}} \Th{m}{N}\left(\tau,\frac{2z}{N}\right)\,
\frac{\th_1(\tau,z)}{i \eta(\tau)^3}.
$$
In the continuous part,   
the `density functions' $\rho_1$ ($\rho_1'$) and $\rho_2$ ($\rho_2'$) 
have rather complicated forms. 
As expected, 
$\rho_1$ ($\rho_1'$) 
includes the logarithmically divergent term as the leading contribution;
$$
\rho_1(p_L,p_R,m;\ep) = C  \left|\ln \ep \right| \delta(p_L-p_R) + \cdots,
$$
with some constant $C$, which corresponds to the strings freely propagating in 
the asymptotic region, and $C\left|\ln \ep \right|$ is roughly identified as 
an infinite volume factor. 
However,  both $\rho_1$ ($\rho_1'$) and 
$\rho_2$ ($\rho_2'$) include subleading, non-diagonal terms with
$p_L \neq p_R$ and considerably non-trivial dependence on $m$, as is mentioned in \cite{ES-NH}.


Now, we would like to  prove the equalities;
\begin{equation}
\rho_1(p_L,p_R,m;\ep) = \rho'_1(p_L,p_R,m;\ep),
\hspace{1cm}
\rho_2(p_L,p_R,m;\ep) = \rho'_2(p_L,p_R,m;\ep).
\label{id rho}
\end{equation}
If this is the case, by using the `charge conjugation relation'\footnote
   {The minus sign just originates from the $\th_1$-factor.}
\begin{equation}
 \chic^{(N,1)}(p,m;\tau,-z) = - \chic^{(N,1)}(p,-m;\tau,z),
\label{cc rel con}
\end{equation}
as well as \eqn{cc rel hchid}, 
we obtain
\begin{eqnarray}
&& Z^{(\stR)}_A(\tau,\bar{\tau}; -z, \bar{z} ; \ep) = 
 e^{-2\pi \hc \frac{z_2^2}{\tau_2}}\, 
\left[ \sum_{v=0}^{N-1} \sum_{a\in \bz_N}\, \hchid^{(N,1)}(N-v,v+a;\tau,z) \,\left[\hchid^{(N,1)}(v,a;\tau,z)\right]^* 
\right.
\nn
&& 
\hspace{1cm}
- \sum_{m\in \bz_{2N}}\, \int dp_L\, \int dp_R\, 
\left\{\rho_1(p_L,p_R, m ; \ep ) \chic^{(N,1)}(p_L, m;\tau,z) 
\left[\chic^{(N,1)}(p_R ,m;\tau,z)\right]^* 
\right.
\nn
&& 
\hspace{3cm}
\left. \left. 
+ \rho_2(p_L,p_R, m ; \ep ) \chic^{(N,1)}(p_L, m;\tau,z) 
\left[\chic^{(N,1)}(p_R ,m+N ;\tau,z)\right]^* \right\}
\right],
\label{ch exp A 2}
\end{eqnarray}
which leads to the desired formula \eqn{part cSL(2)/U(1)}.


~

\noindent
{\bf [proof of \eqn{id rho}]}

We start with  recalling  the `orbifold relation' between   $Z^{(\stR)}_A$ and $Z^{(\stR)}_V$, which is  shown in \cite{ES-NH,orb-ncpart};
\begin{eqnarray}
Z^{(\stR)}_{A}(\tau,\bar{\tau}, z , \bar{z} ;\ep) &=& \frac{e^{\frac{2\pi}{\tau_2} \hc \left|z\right|^2}}{N}\,
 \sum_{\al_1, \al_2 \in \bz_N}\, Z_{V, (\al_1,\al_2)}(\tau,\bar{\tau}; z , \bar{z} ;\ep),
\label{orbifold relation}
\end{eqnarray}
where we introduced the `twisted partition function'\footnote
  {The relation of the notations here and given in  \cite{orb-ncpart} is as follows; 
\begin{eqnarray*}
&& Z^{(\stR)}_A(\tau,\bar{\tau}; z, \bar{z};\ep)= Z_{\reg}(\tau,z;\ep),  \hspace{1cm}
Z^{(\stR)}_V(\tau,\bar{\tau}; z, \bar{z};\ep) = \tZ_{\reg}(\tau,z ;\ep), 
, \hspace{1cm}
\\
&&
Z^{(\stR)}_{V, \, (\beta_1, \beta_2)}(\tau,\bar{\tau}; z, \bar{z};\ep) =
\tZ^{(N)}_{\reg}(\tau,z | \beta_1, \beta_2 ;\ep) .
\end{eqnarray*}
}
in the R.H.S of \eqn{orbifold relation};
\begin{eqnarray}
\hspace{-1cm}
Z^{(\stR)}_{V, (\al_1,\al_2)}(\tau,\bar{\tau}; z , \bar{z} ;\ep) &=& 
N  e^{- \frac{2\pi}{\tau_2} \frac{N+4}{N} z_2^2}\,
\sum_{n_1,n_2 \in \bz}\,
\int_{\Sigma(z,\ep)} \frac{d^2u}{\tau_2}\, 
 e^{- 4\pi \frac{u_2 z_2}{\tau_2}}\,
\left|\frac{\th_1\left(\tau, u + \frac{N+2}{N}z\right)}
{\th_1\left(\tau, u + \frac{2}{N} z \right)} \right|^2 
\nn
&& \hspace{1cm} \times 
 e^{-\frac{\pi}{k \tau_2}\left| (N n_1 +\al_1) \tau+ (N n_2 +\al_2) \right|^2}
\, e^{2\pi i \left\{  (N n_2 + \al_2) s_1 - (N n_1 +\al_1 ) s_2  \right\}}.
\label{ZV twisted}
\end{eqnarray}
%
We can rewrite \eqn{ZV twisted} by means of the Poisson resummation;
\begin{eqnarray}
Z^{(\stR)}_{V, (\al_1,\al_2)}(\tau,\bar{\tau}; z , \bar{z} ;\ep)
&=&
e^{-2\pi \frac{\hc}{\tau_2}z_2^2} \,
\left|\frac{\th_1(\tau,z)}{\eta(\tau)^3}\right|^2 \,
\sum_{v\in \bz}\, \sum_{\stackrel{\ell,\tell\in \bz}{\ell-\tell \in \al_1 + N\bz}}\, e^{2\pi i \frac{\al_2}{N}(v+\ell+\tell)}\, 
\nn
&&
\hspace{1cm} \times
\frac{1}{2\pi i} \, \left\lb
\int_{\br-i0} dp\, 
e^{-\vep (v+ip)} yq^{\ell} \left[yq^{\tell} \right]^* - \int_{\br+i(N-0)} dp\, e^{\vep (v+ip)}
\right\rb
\nn
&& \hspace{1cm} \times \, \frac{e^{-\pi \tau_2 \frac{p^2+v^2}{N}}} {p-iv}
\,
\frac{(yq^{\ell})^{\frac{v}{N}}}{1-yq^{\ell}} \,
\left[
\frac{(yq^{\tell})^{\frac{v}{N}}}{1-yq^{\tell}}
\right]^*\,
y^{\frac{2\ell}{N}} q^{\frac{\ell^2}{N}} \, 
\left[y^{\frac{2\tell}{N}} q^{\frac{\tell^2}{N}}\right]^*,
\label{ZV twisted 2}
\end{eqnarray}
where we set $\vep := 2\pi \frac{\tau_2}{N}\ep$.
A closely related analysis is given in \cite{ES-NH}. 
We obviously find 
$$
Z^{(\stR)}_{V, (0,0)}(\tau,\bar{\tau}, z , \bar{z} ;\ep) = Z^{(\stR)}_{V}(\tau,\bar{\tau}, z , \bar{z} ;\ep),
$$
and \eqn{ZV twisted 2} immediately implies the relation
\begin{equation}
Z^{(\stR)}_{V, (\al_1,\al_2)}(\tau,\bar{\tau}; z , \bar{z} ;\ep) 
= s^L_{(\al_1,\al_2)} \cdot Z^{(\stR)}_V(\tau,\bar{\tau}; z, \bar{z} ;\ep),
\end{equation}
where $s^L_{(\al_1, \al_2)}$ $(\al_i \in \bz)$ 
denotes the spectral flow operator  \eqn{def s ab} acting only on the left-mover;
\begin{equation}
s^L_{(\al_1,\al_2)}\cdot f(\tau,\bar{\tau}, z, \bar{z}) := (-1)^{\al_1+\al_2} q^{\frac{\hc}{2} \al_1^2} y^{\hc \al_1} e^{2\pi i \frac{\al_1 \al_2}{N}}\,
 f(\tau,\bar{\tau}, z+\al_1 \tau+\al_2, \bar{z})
\end{equation}
Finally, by using the identity
\begin{eqnarray}
\frac{1}{N}\sum_{\al_1,\al_2\in \bz_N}\, s_{(\al_1,\al_2)}\cdot 
\hchid^{(N,1)}(v,a;\tau,z)
&=& \hchid^{(N,1)}(v,-v-a;\tau,z),
\label{orb rel chid}
\\
\frac{1}{N}\sum_{\al_1,\al_2\in \bz_N}\, s_{(\al_1,\al_2)}\cdot 
\chic^{(N,1)}(p,m;\tau,z)
&=& \chic^{(N,1)}(p,-m;\tau,z),
\label{orb rel chic}
\end{eqnarray}
which are easily proven by direct calculations\footnote
  {\eqn{orb rel chic} is essentially the same identity as \eqn{orbifolding theta}.
}, we can achieve the wanted identities
\eqn{id rho}. 
\hspace{5mm}
{\bf (Q.E.D)}

~


\subsection{Other Spin Structures}

We finally briefly discuss about other spin structures. 
We shall assume the diagonal spin structures and the non-chiral GSO projection.

We start with the ansatz of the total partition function; 
\begin{eqnarray}
Z(\tau,\bar{\tau}; z, \bar{z}) &=& \frac{1}{2} \sum_{\sigma=\sNS,\stNS,\sR,\stR}\, \left[
Z_V^{(\sigma)}(\tau,\bar{\tau}; z, \bar{z}) + \vep(\sigma) Z_A^{(\sigma)}(\tau,\bar{\tau}; - z, \bar{z})
\right],
\label{ansatz Z}
\end{eqnarray}
where the partition functions with left-right symmetric spin structures $\sigma = \NS, \tNS, \R, \tR$ are
evaluated by means of path-integration as in \eqn{part fn V} and \eqn{part fn A}. 
Relevant calculations for general spin structures as well as helpful formulas are presented 
in \cite{Sug-NENS5}\footnote
   {Since we are now assuming the non-chiral GSO projection, we do not need choose the parameter as  $\hc = 1 + \frac{2K}{N}$, $K \in 2\bz_{>0}$ 
($N$ and $K$ are not necessarily co-prime) as in \cite{Sug-NENS5}. 
This assumption is necessary when considering the {\em chiral\/} GSO projection.
}.

We set the sign factor $\vep(\tR)=1$ to reproduce \eqn{part cSL(2)/U(1)} when setting $\sigma = \tR$, 
and we have the following two possibilities of $\vep(\sigma)$ for other spin structures 
that are compatible with the modular invariance;


\begin{description}
\item[(i) $\vep(\sigma)=1$, $\any \sigma$ :   `non-geometric deformation of $SL(2)/U(1)$-supercoset' ]

~

With this naive choice, the continuous parts of partition functions of the vector and axial types  
are common and appear {\em with the same sign\/} contrary to the $\tR$-case. 
This feature just originates from the simple fact; $\th_3(z)$, $\th_4(z)$, $\th_2(z)$ are even functions of $z$
although $\th_1(z)$ is an odd function. 
Thus, there are continuous excitations in the sectors $\sigma=\NS,\tNS, \tR$ as in the standard $SL(2)/U(1)$-supercoset.

On the other hand, the discrete part of each spin structure is described by the following building blocks 
($v\in \bz_N$, $a\in \frac{1}{2}+\bz_N$ for $\sigma= \NS, \tNS$, $a\in \bz_N$ for $\sigma = \R,\tR$);
\begin{eqnarray}
X_+^{(N)\,[\sNS]}(v,a;\tau,z) &:= & \hchid^{(N,1)\, [\sNS]}(v,a;\tau,z) + \hchid^{(N,1)\, [\sNS]}(N-v, a+v;\tau,z), 
\nn
X_-^{(N)\,[\stNS]}(v,a;\tau,z) &:= & \hchid^{(N,1)\, [\stNS]}(v,a;\tau,z) - \hchid^{(N,1)\, [\stNS]}(N-v, a+v;\tau,z), 
\nn
X_+^{(N)\,[\sR]}(v,a;\tau,z) &:= & \hchid^{(N,1)\, [\sR]}(v,a;\tau,z) + \hchid^{(N,1)\, [\sR]}(N-v, a+v;\tau,z), 
\nn
X_+^{(N)\,[\stR]}(v,a;\tau,z) &:= & \hchid^{(N,1)\, [\stR]}(v,a;\tau,z) + \hchid^{(N,1)\, [\stR]}(N-v, a+v;\tau,z)
\left(\equiv \eqn{X hchid}\right), 
\nn
\label{X spin str 1}
\end{eqnarray}
where we have explicitly indicated the spin structure.  
The explicit definitions of modular completions with general spin structures are presented in  \cite{Sug-NENS5}.
We remark that the functions $X_*^{(N)\,[\sigma]}(v,a)$ appearing in \eqn{X spin str 1}
are generically  {\em non-holomorphic\/} except for the $\tR$-sector.

This model shares the asymptotic cylindrical region with the radius $\frac{1}{N} \sqrt{N\al'} \equiv \sqrt{\frac{1}{N}\al'}$
with the standard $SL(2)/U(1)$ (of the vector type)
and aspect of propagating strings is almost the same. 
However, we have non-trivial deformations in the discrete spectrum, which 
leads us to the holomorphic elliptic genus \eqn{EG cSL(2)/U(1)} and 
would be {\em non-geometric\/} 
since they are never realized only within the cigar theory (or the trumpet theory).


~

\item[(ii) $\vep(\sigma)=-1$ for $\sigma=\NS,\tNS,\R$, and $\vep(\tR)=1$ : ]
{\bf `Compactified $SL(2)/U(1)$-Supercoset' }

This  second possibility is more curious. In this case, the continuous sectors 
are canceled out for all the spin structures, and the discrete parts are
described  respectively by 
\begin{eqnarray}
X_-^{(N)\,[\sNS]}(v,a;\tau,z) &:= & \hchid^{(N,1)\, [\sNS]}(v,a;\tau,z) - \hchid^{(N,1)\, [\sNS]}(N-v, a+v;\tau,z), 
\nn
X_+^{(N)\,[\stNS]}(v,a;\tau,z) &:= & \hchid^{(N,1)\, [\stNS]}(v,a;\tau,z) + \hchid^{(N,1)\, [\stNS]}(N-v, a+v;\tau,z), 
\nn
X_-^{(N)\,[\sR]}(v,a;\tau,z) &:= & \hchid^{(N,1)\, [\sR]}(v,a;\tau,z) - \hchid^{(N,1)\, [\sR]}(N-v, a+v;\tau,z), 
\nn
X_+^{(N)\,[\stR]}(v,a;\tau,z) &:= & \hchid^{(N,1)\, [\stR]}(v,a;\tau,z) + \hchid^{(N,1)\, [\stR]}(N-v, a+v;\tau,z)
\left(\equiv \eqn{X hchid}\right).
\nn
\label{X spin str 2}
\end{eqnarray}
Namely, we achieve the total partition function in a very simple form\footnote
   {The overall factor $1/4$ is naturally interpreted as follows;  
 $1/2$ originates from the non-chiral GSO projection, while the remaining $1/2$ 
can avoid the over counting due to the obvious $\bz_2$-symmetry 
$$
X_{\pm}^{(N)\,[\sigma]}(v,a) = \pm X_{\pm}^{(N)\,[\sigma]}(N-v,v+a).
$$ 
}; 
\begin{eqnarray}
&& \hspace{-1cm}
Z_{cSL(2)/U(1)}(\tau,\bar{\tau};z,\bar{z}) =
 \frac{1}{4} e^{-\frac{2\pi}{\tau_2}\hc z_2^2}\, \sum_{v\in \bz_N}\, \sum_{a\in \bz_N} \,
\left[  
 \left|X_-^{(N)\,[\sNS]}(v,a+\frac{1}{2};\tau,z)\right|^2  \right.
 \nn
 && 
 \left. +  \left|X_+^{(N)\,[\sNS]}(v,a+\frac{1}{2};\tau,z)\right|^2
+ \left|X_-^{(N)\,[\sR]}(v,a;\tau,z)\right|^2 + \left|X_+^{(N)\,[\stR]}(v,a;\tau,z)\right|^2
\right].
\label{part cSL(2)/U(1) total}
\end{eqnarray}
This is a natural extension of \eqn{part cSL(2)/U(1)} including all the spin structures,  
and should be directly compared with the $\cN=2$ minimal model (with level $N-2$);
\begin{equation}
Z_{\msc{min}}(\tau,\bar{\tau};z,\bar{z}) = \frac{1}{4} e^{-\frac{2\pi}{\tau_2}\left(1-\frac{2}{N}\right) z_2^2}\, 
\sum_{\sigma=\sNS,\stNS,\sR,\stR} \, \sum_{\ell=0}^{N-2}\, \sum_{m\in \bz_N}\, 
\left|\ch{(\sigma)}{\ell,m} (\tau,z) \right|^2.
\label{part minimal}
\end{equation}

All of the building blocks $X_*^{(N) \, [\sigma]}(v,a)$ in \eqn{part cSL(2)/U(1) total} are 
holomorphic and can be rewritten in terms only of the extended characters (that are {\em not\/} modular completed) as in 
\eqn{X chid}.
In fact, the functions \eqn{X spin str 2} for $\sigma= \NS, \tNS, \R$ are reproduced by the `half spectral flows' ;
$z\,\mapsto\, z+ \frac{\tau+1}{2}$,  $z\,\mapsto\, z+ \frac{\tau}{2}$, $z\,\mapsto\, z+ \frac{1}{2}$.  
The absence of non-holomorphic corrections means that they are directly associated to some infinitely reducible representations 
of $\cN=2$ SCA with $\hc=1+\frac{2}{N}$. These representations are, however, {\em non-unitary\/}  due to the relative minus sign 
appearing in the NS and R-sectors. 
This aspect is in a sharp contrast to the $\cN=2$ minimal model, in which the  partition function \eqn{part minimal} 
only includes the characters of unitary irreducible representations.

\end{description}

~


\section{Summary and Comments}

In this paper,  we have studied a possible `analytic continuation' with $N \rightarrow -N$
of the ${\cal N}=2$ minimal model with the central charge $\hat{c} = 1 - \frac{2}{N}$.
Namely, we have examined the problem of what is the superconformal system with $\hat{c} = 1 + \frac{2}{N}$ that has 
\eqn{Z expected} as its elliptic genus.

Our main results are summarized as follows;
\begin{description}
\item[(i)]

The `Fourier expansion' of the function \eqn{Z expected} is 
rewritten by a {\em holomorphic\/} linear combination 
of the modular completions of the extended discrete characters of $SL(2)/U(1)$-model \cite{ES-NH}.
This result is exhibited 
in terms of the formula \eqn{X hchid} or  \eqn{Z tZ Phi}, equivalently. 
This is similar to the fact that the elliptic genus of
$\cN=2$ minimal model $\cZ^{(N-2)}_{\msc{min}}(\tau,z)$ \eqn{Z min}
is Fourier expanded by the characters associated to the Ramond ground states \cite{Witten-EG minimal}.

\item[(ii)] 

The superconformal system corresponding to \eqn{Z expected} is identified with 
a {\em `compactified'} model of $SL(2)/U(1)$-supercoset, as is given by 
\eqn{part cSL(2)/U(1)}.

\item[(iii)]

Two possibilities of extending to 
general spin structures have been presented;
One is a non-compact model regarded as 
a `non-geometric deformation' of $SL(2)/U(1)$-supercoset,
and the other is the natural extension 
of the compactified model \eqn{part cSL(2)/U(1)}. 
The latter is quite similar to the $\cN=2$ minimal model, 
although it is not a unitary theory.

\end{description}


~

We would like to add a few comments;





The partition function \eqn{part cSL(2)/U(1)} (and \eqn{part cSL(2)/U(1) total}) looks very like those of RCFTs. 
We only possess finite conformal blocks that are holomorphically factorized in the usual sense.
However, there is a crucial difference from generic RCFTs defined axiomatically.  
The partition function \eqn{part cSL(2)/U(1)} or the elliptic genus \eqn{EG cSL(2)/U(1)} 
does not include the contributions from the Ramond vacua saturating the unitarity bound $Q = \pm \frac{\hc}{2}$.
This implies that the Hilbert space of normalizable states does not contain the NS-vacuum ($h=Q=0$)
which should correspond to the identity operator. 
Of course, this feature is common with  the spectrum of original $SL(2)/U(1)$-supercoset 
read off from the torus partition function evaluated in \cite{ES-BH} (see also \cite{IKPT,HPT,LST}). 
It may be an interesting question whether or not the finiteness of conformal blocks {\em without the identity representation\/},
which is observed in our `compactified $SL(2)/U(1)$-model', 
unavoidably leads to a non-unitarity of the spectrum in general conformal field theories.





A natural extension of this work would be the study of the cases of `fractional levels'
$ \hc = 1 + \frac{2K}{N}$, $(K\geq 2, ~ \mbox{GCD}\{N,K\}=1)$.  
In other words, one may search a theory of which elliptic genus would be 
$$
\cZ(\tau,z) = K \Phi^{(N/K)}(\tau,z) \equiv K \frac{\th_1\left(\tau, \frac{N+K}{N}z\right)}{\th_1\left(\tau, \frac{K}{N}z\right)},
$$
which has the Witten index $\cZ(\tau, z=0) = N+K$. 
However, the function $\Phi^{(N/K)}(\tau,z)$ is only {\em meromorphic\/} with respect to the angle variable $z$, 
and such a function is not likely to be realized as the elliptic genus of 
any superconformal field theory.

We also point out that the cancellation of continuous parts such as \eqn{part cSL(2)/U(1)} does not seem to happen 
in that case. This fact suggests that the `compactification' of $SL(2)/U(1)$-supercoset works only for integer levels, that is,  
$\hc= 1+ \frac{2}{N}$.


~


\section*{Acknowledgments}

The author should thank T. Eguchi for useful discussions
at the early stage of this work.

This research was supported by JSPS KAKENHI Grant Number 23540322 
from Japan Society for the Promotion of Science (JSPS).





\newpage

\section*{Appendix A: ~ Conventions for Theta Functions}

\setcounter{equation}{0}
\def\theequation{A.\arabic{equation}}

We assume $\tau\equiv \tau_1+i\tau_2$, $\tau_2>0$ and 
 set $q:= e^{2\pi i \tau}$, $y:=e^{2\pi i z}$;
 \begin{equation}
 \begin{array}{l}
 \dsp \th_1(\tau,z)=i\sum_{n=-\infty}^{\infty}(-1)^n q^{(n-1/2)^2/2} y^{n-1/2}
  \equiv 2 \sin(\pi z)q^{1/8}\prod_{m=1}^{\infty}
    (1-q^m)(1-yq^m)(1-y^{-1}q^m), \\
 \dsp \th_2(\tau,z)=\sum_{n=-\infty}^{\infty} q^{(n-1/2)^2/2} y^{n-1/2}
  \equiv 2 \cos(\pi z)q^{1/8}\prod_{m=1}^{\infty}
    (1-q^m)(1+yq^m)(1+y^{-1}q^m), \\
 \dsp \th_3(\tau,z)=\sum_{n=-\infty}^{\infty} q^{n^2/2} y^{n}
  \equiv \prod_{m=1}^{\infty}
    (1-q^m)(1+yq^{m-1/2})(1+y^{-1}q^{m-1/2}), \\
 \dsp \th_4(\tau,z)=\sum_{n=-\infty}^{\infty}(-1)^n q^{n^2/2} y^{n}
  \equiv \prod_{m=1}^{\infty}
    (1-q^m)(1-yq^{m-1/2})(1-y^{-1}q^{m-1/2}) .
 \end{array}
\label{th}
 \end{equation}
 \begin{eqnarray}
 \Th{m}{k}(\tau,z)&=&\sum_{n=-\infty}^{\infty}
 q^{k(n+\frac{m}{2k})^2}y^{k(n+\frac{m}{2k})} .
 \end{eqnarray}
 We also set
 \begin{equation}
 \eta(\tau)=q^{1/24}\prod_{n=1}^{\infty}(1-q^n).
 \end{equation}
%
%
The spectral flow properties of theta functions are summarized 
as follows ($m,n, a \in \bz$, $k \in \bz_{>0}$);
\begin{eqnarray}
 && \th_1(\tau, z+m\tau+n) = (-1)^{m+n} 
q^{-\frac{m^2}{2}} y^{-m} \th_1(\tau,z) ~, \nn
&& \th_2(\tau, z+m\tau+n) = (-1)^{n} 
q^{-\frac{m^2}{2}} y^{-m} \th_2(\tau,z) ~, \nn
&& \th_3(\tau, z+m\tau+n) = 
q^{-\frac{m^2}{2}} y^{-m} \th_3(\tau,z) ~, \nn
&& \th_4(\tau, z+m\tau+n) = (-1)^{m} 
q^{-\frac{m^2}{2}} y^{-m} \th_4(\tau,z) ~, \nn
&& \Th{a}{k}(\tau, 2(z+m\tau+n)) = 
q^{-k m^2} y^{-2 k m} \Th{a+2km}{k}(\tau,2z)~.
\label{sflow theta}
\end{eqnarray}

%

~


\section*{Appendix B:~ Summary of Modular Completions}

\setcounter{equation}{0}
\def\theequation{B.\arabic{equation}}


In this appendix we summarize the definitions as well as 
useful formulas for the `extended discrete characters'  
and their modular completions of the $\cN=2$ superconformal algebra with 
$\hc \left(\equiv \frac{c}{3} \right)= 1+ \frac{2}{k}$.
We focus only on  the $\tR$-sector\footnote
   {In this paper we shall use the convention of $\tR$-characters 
with the same sign as \cite{orb-ncpart}, and the inverse sign compared to those of \cite{ES-NH,ES-BH,ES-C}, 
so that the Witten indices appear with the positive sign. 
(See \eqn{IR chid} below.)}, and when treating the extended characters, 
we assume $k= N/K$, $(N,K \in \bz_{>0}) $ (but, not assume $N$ and $K$ are co-prime).

~

\noindent
{\bf \underline{Extended Continuous (non-BPS) Characters \cite{ES-L,ES-BH}:}}
\begin{eqnarray}
\chic^{(N,K)} (p,m;\tau,z) &:=  & 
q^{\frac{p^2}{4NK}} \Th{m}{NK}\left(\tau,\frac{2z}{N}\right)\,
\frac{\th_1(\tau,z)}{i \eta(\tau)^3}.
\label{chic}
\end{eqnarray}
This corresponds to the spectral flow sum of the non-degenerate representation with
$h= \frac{p^2+m^2}{4NK} + \frac{\hc}{8}$, 
$Q = \frac{m}{N}\pm \frac{1}{2}$~($p\geq 0$, $m\in \bz_{2NK}$),
whose flow momenta are taken to be $n\in N \bz$.
The modular and spectral flow properties are simply written as 
\begin{eqnarray}
&& 
\hspace{-1cm}
\chic^{(N,K)}\left(p,m ; - \frac{1}{\tau}, \frac{z}{\tau}\right)
= (-i) e^{i\pi \frac{\hc}{\tau}z^2}\, 
 \frac{1}{2NK} \int_{-\infty}^{\infty} dp' \, 
\sum_{m'\in\bz_{2NK}}\, 
e^{2\pi i \frac{p p' - m m'}{2NK}}
\, \chic^{(N,K)} (p',m';\tau,z) .
\label{S chic}
\\
&& \hspace{-1cm}
\chic^{(N,K)}\left(p,m ; \tau+1, z \right)
= e^{2\pi i \frac{p^2+m^2}{4NK}}\,
\chic^{(N,K)} \left(p, m ; \tau, z \right),
\label{T chic}
\\
&& \hspace{-1cm}
\chic^{(N,K)} (p,m;\tau,z+r\tau+s) = (-1)^{r+s} e^{2\pi i \frac{m}{N}s}
q^{-\frac{\hc}{2}r^2} y^{-\hc r}\, \chic^{(N,K)}(p,m+2Kr;\tau,z),
~~~ (\any r,s \in \bz).
\nn
&&
\label{sflow chic}
\end{eqnarray}

~

\noindent
{\bf \underline{Extended Discrete Characters \cite{ES-L,ES-BH,ES-C}:}}
\begin{eqnarray}
\chid^{(N,K)} (v,a;\tau,z) &:= & 
\sum_{n\in a + N\bz}\,  \frac{(yq^{n})^{\frac{v}{N}}}
{1-yq^{n}} \, y^{\frac{2K}{N}n}q^{\frac{K}{N} n^2}\, \frac{\th_1(\tau,z)}{i \eta(\tau)^3}\,, 
\label{chid}
\end{eqnarray}
This again corresponds to the  sum of the 
Ramond vacuum representation with $h= \frac{\hc}{8}$, $Q= \frac{v}{N}-\frac{1}{2}$
~($v=0,1,\ldots , N$) over spectral flow
 with flow momentum $m$  taken to be mod.$N$, as 
$m= a +N\bz$ ~ ($a\in \bz_N$).
If one introduces the notation of Appell function or Lerch sum with level $2k$
\cite{Pol,STT,Zwegers}, 
\begin{equation}
 \cK^{(2k)}(\tau,z) 
:= \sum_{n\in \bsz} \frac{y^{2kn} q^{kn^2} }
{1-yq^n},
\label{Appell}
\end{equation}
$\chid(v,a)$ is identified as its Fourier expansion;
\begin{eqnarray}
&& y^{\frac{2K}{N} a} q^{\frac{K}{N} a^2} \,
 \cK^{(2NK)}\left(\tau, \frac{z+a\tau+b}{N}\right)\, 
\frac{\th_1(\tau,z)}{i \eta(\tau)^3} 
= \sum_{v=0}^{N-1} \,  e^{2\pi i \frac{v b}{N}} \, \chid^{(N,K)} (v,a;\tau,z).
\label{rel chid cK} 
\end{eqnarray}
We also note 
\begin{equation}
\chid^{(N,K)}(N,a;\tau,z) = \chid^{(N,K)}(0,a;\tau,z) - \Th{2Ka}{NK}\left(\tau, \frac{2z}{N} \right) \frac{\th_1(\tau,z)}{i \eta(\tau)^3} .
\label{rel chid N 0}
\end{equation}

~


The modular transformation formulas of
$\chid^{(N,K)}(v,a)$
and $\cK^{(2k)}$ 
can be expressed as \cite{ES-L,ES-BH,ES-C,STT,Zwegers};
\begin{eqnarray}
&& 
\hspace{-1cm}
\chid^{(N,K)} \left(v,a ; - \frac{1}{\tau}, \frac{z}{\tau}\right)
= e^{i\pi \frac{\hc}{\tau}z^2}\,\left[
 \sum_{v=0}^{N-1} \,\sum_{a\in \bz_N}\,
\frac{1}{N} \, e^{2\pi i \frac{vv' - (v+2Ka)(v'+2Ka')}{2NK}}
\, \chid^{(N,K)}(v',a';\tau,z) \right.
\nn
&& \hspace{1.5cm}
\left. -\frac{i}{2NK} \sum_{m' \in \bz_{2NK}} \, e^{-2\pi i \frac{(v+2Ka) m'}{2NK}}\,
\int_{\br+i0} dp'\, \frac{e^{-2\pi \frac{vp'}{2NK}}}{1-e^{-\pi \frac{p'+im'}{K}}}
\, \chic^{(N,K)} (p',m';\tau,z)
\right],
\label{S chid}
\\
&& 
\hspace{-1cm}
\chid^{(N,K)} \left(v,a ; \tau+1, z \right)
= e^{2\pi i \frac{a}{N} \left(v+ K a \right)}\,
\chid^{(N,K)} \left(v,a ; \tau, z \right),
\label{T chid} 
\end{eqnarray}
\begin{eqnarray}
&&\cK^{(2 k)}\left(-\frac{1}{\tau}, \frac{z}{\tau}\right)
= \tau e^{i\pi  \frac{ 2k z^2}{\tau}}\,
\left[ \cK^{(2k)}(\tau,z) - \frac{i}{\sqrt{2 k}}\, \sum_{m\in \bz_{2k}}\,
\int_{\br+i0} dp' \, \frac{q^{\frac{1}{2}p^{'2}}}
{1-e^{-2\pi \left(\frac{p'}{\sqrt{2k}}+i\frac{m}{2k}\right)}}\,
\Th{m}{k}(\tau,2z)
\right],
\nn
&&
\label{S cK}
\\
&& \cK^{(2k)} (\tau+1, z) = \cK^{(2k)}(\tau,z).
\label{T cK}
\end{eqnarray}
%

The spectral flow property is also expressed as 
\begin{equation}
\chid^{(N,K)} (v,a;\tau,z+r\tau+s) = (-1)^{r+s} e^{2\pi i \frac{v+2Ka}{N}s}
q^{-\frac{\hc}{2}r^2} y^{-\hc r}\, \chid^{(N,K)}(v,a+r;\tau,z),
~~~ (\any r,s \in \bz),
\label{sflow chid}
\end{equation}
\begin{equation}
\cK^{(2k)} (\tau, z+ r\tau+s) = q^{-k r^2} y^{-2k r} \cK^{(2k)}(\tau,z), ~~~ (\any r,s \in \bz),
\label{sflow cK}
\end{equation}

~

\noindent
{\bf \underline{Modular Completion of the Extended Discrete Characters:}}

The modular completion of the discrete character $\chid$ is defined as follows;
\begin{eqnarray}
\hspace{-1cm}
\hchid^{(N,K)} (v,a;\tau,z) 
&: =& 
\frac{\th_1(\tau,z)}{ 2\pi \eta(\tau)^3}\,
\sum_{\stackrel{n\in a + N\bz}{r \in v + N\bz}}\, 
\left\{ \int_{\br + i(N-0)} dp\, -
\int_{\br-i0} dp \, \left(y q^{n} \right) 
\right\}
\,
\frac{ e^{- \pi \tau_2 \frac{p^2+r^2}{NK}} 
\left(y q^{n}\right)^{\frac{r}{N}}}{p-ir} 
\, \frac{y^{\frac{2K}{N}n} q^{\frac{K}{N}n^2}}{1-y q^{n}}
\nn
&=& \chid^{(N,K)}(v,a;\tau,z) 
+ \frac{\th_1(\tau,z)}{ 2\pi \eta(\tau)^3}\,
\sum_{\stackrel{n\in a + N\bz}{r \in v + N\bz}}\, 
\int_{\br -i 0} dp \,
\frac{ e^{- \pi \tau_2 \frac{p^2+r^2}{NK}} }{p-ir} 
\, \left(y q^{n}\right)^{\frac{r}{N}} y^{\frac{2K}{N}n} q^{\frac{K}{N}n^2}.
\nn
&&
\label{hchid}
\end{eqnarray}


This expression \eqn{hchid} is obviously periodic with respect to both of $v$ and $a$;
\begin{equation}
\hchid^{(N,K)}(v+Nm, a+Nn;\tau,z) = \hchid^{(N,K)} (v,a;\tau,z), \hspace{1cm} (\any m,n \in \bz).
\label{periodicity hchid}
\end{equation}
Especially, we have 
\begin{equation}
\hchid^{(N,K)}(N,a;\tau,z) = \hchid^{(N,K)}(0,a;\tau,z),
\label{rel hchid N 0}
\end{equation}
in spite of \eqn{rel chid N 0}.

~


The modular completion of Appell function $\cK^{(2k)}(\tau,z)$ is 
given as \cite{Zwegers};
\begin{eqnarray}
\hcK^{(2k)} (\tau, z) &:=& \cK^{(2k)}(\tau,z) - \frac{1}{2} \sum_{m\in \bz_{2k}}\,
R_{m,k}(\tau) \, \Th{m}{k}(\tau,2z),
\label{hcK}
\end{eqnarray}
where we set 
\begin{eqnarray}
R_{m,k}(\tau) &:=& 
\frac{1}{i\pi}\, \sum_{r \in m+2k\bz}\,
\int_{\br- i0} dp \, \frac{e^{-\pi \tau_2 \frac{p^2+r^2}{k}} }{p-ir}\,
q^{- \frac{r^2}{4k}},
\label{Rmk}
\end{eqnarray}
which is generically non-holomorphic due to the $\tau_2$-dependence.

One can easily show 
\begin{eqnarray}
&& R_{m+2k s,k} (\tau) = R_{m,k}(\tau), ~~~ (\any s \in \bz), \hspace{1cm} 
R_{m,k} (\tau)= 2 \delta^{(2k)}_{m,0} - R_{- m,k} (\tau) ,
\label{R identity}
\end{eqnarray}
and thus 
\begin{equation}
R_{0,k}(\tau) \equiv 1 , \hspace{1cm} R_{k,k}(\tau) \equiv 0,
\label{R identity 2}
\end{equation}
holds, especially.


The `Fourier expansion relation' \eqn{rel chid cK} is inherited to the modular completions;
\begin{eqnarray}
&& y^{\frac{2K}{N} a} q^{\frac{K}{N} a^2} \,
 \hcK^{(2NK)}\left(\tau, \frac{z+a\tau+b}{N}\right)\, 
\frac{\th_1(\tau,z)}{i \eta(\tau)^3} 
= \sum_{v\in \bz_N} \,  e^{2\pi i \frac{v b}{N}} \, \hchid^{(N,K)} (v,a;\tau,z).
\label{rel hchid hcK} 
\end{eqnarray}


%
%

The modular transformation formulas for the modular completions
\eqn{hchid}, \eqn{hcK} are written as 
\begin{eqnarray}
&& \hspace{-1cm}
\hchid^{(N,K)} \left(v,a ; - \frac{1}{\tau}, \frac{z}{\tau}\right)
= e^{i\pi \frac{\hc}{\tau}z^2}\, \sum_{v'=0}^{N-1} \,\sum_{a'\in \bz_N}\,
\frac{1}{N} \, e^{2\pi i \frac{vv' - (v+2Ka)(v'+2Ka')}{2NK}}
\, \hchid^{(N,K)} (v',a';\tau,z),
\label{S hchid}
\\
&& \hspace{-1cm}
\hchid^{(N,K)} \left(v,a ; \tau+1, z \right)
= e^{2\pi i \frac{a}{N} \left(v+ K a \right)}\,
\hchid^{(N,K)} \left(v,a ; \tau, z \right).
\label{T hchid}
\\
&& 
\hcK^{(2k)} \left(-\frac{1}{\tau}, \frac{z}{\tau} \right) = \tau e^{i\pi \frac{2k}{\tau}z^2}\, \hcK^{(2k)}(\tau,z), 
\label{S hcK}
\\
&& 
\hcK^{(2k)} (\tau+1, z) = \hcK^{(2k)} (\tau, z).
\label{T hck}
\end{eqnarray}
When compared with \eqn{S chid} and  \eqn{S cK}, 
the S-transformation formulas have been simplified because of the absence of continuous terms.

Also the spectral flow property is preserved by taking the completion;
\begin{equation}
\hchid^{(N,K)} (v,a;\tau,z+r\tau+s) = (-1)^{r+s} e^{2\pi i \frac{v+2Ka}{N}s}
q^{-\frac{\hc}{2}r^2} y^{-\hc r}\, \hchid^{(N,K)}(v,a+r;\tau,z),
~~~ (\any r,s \in \bz).
\label{sflow hchid}
\end{equation}
\begin{equation}
\hcK^{(2k)} (\tau, z+ r\tau+s) = q^{-k r^2} y^{-2k r} \hcK^{(2k)}(\tau,z),
~~~ (\any r,s \in \bz).
\label{sflow hcK}
\end{equation}

~


Note that the IR-part of modular completions are evaluated as 
\begin{equation}
\left[ \hchid^{(N,K)} (v,a;\tau,z) \right] \left(\equiv \lim_{\tau_2\rightarrow \infty}\, \hchid^{(N,K)} (v,a;\tau,z) \right)
= \left\{
\begin{array}{ll}
\dsp   \delta_{a,0}^{(N)} \, \frac{1}{2}\left(y^{-\frac{1}{2}} + y^{\frac{1}{2}}\right) & ~~ (v=0, N) \\
\dsp  \delta_{a,0}^{(N)} \, y^{\frac{v}{N} -\frac{1}{2}} & ~~ (v=1,\ldots, N-1),
\end{array}
\right.
\label{IR hchid}
\end{equation}
while we have 
\begin{equation}
\left[ \chid^{(N,K)} (v,a;\tau,z) \right]  = \delta_{a,0}^{(N)} \, y^{\frac{v}{N} -\frac{1}{2}}, \hspace{1cm} (v=0,\ldots, N).
\label{IR chid}
\end{equation}

~


\section*{Appendix C:~ Finiteness of $R_{a,b}(\tau)$ at the cusp $\tau=i\infty$}

\setcounter{equation}{0}
\def\theequation{C.\arabic{equation}}


In Appendix C, we confirm the finiteness of the residue function $R_{a,b}(\tau)$
\eqn{Rab} at the cusp $\tau=i\infty$.
Namely we will prove
\begin{equation}
\lim_{\tau  \rightarrow  i \infty} \left| R_{a,b}(\tau) \right| < \infty,
\label{Rab cusp 2}
\end{equation}
for $\any (a,b) \in \bz_{N+1}\times \bz_{N+1} -  \{ (0,0) \}$.


We classify $z_{a,b}$ \eqn{z ab}  into three groups and examine the behavior of $F(\tau,z)$ \eqn{def F} around them separately;
\begin{description}
\item[(i) $a=0, ~ b=1, \ldots, N$ :]

~

In this case, we have  $\dsp z_{0,b} \equiv \frac{b}{N+1}$,  ($b=1,\ldots, N$).
All of them are simple zeros of the function;
$$
\left[\Phi^{(N)}(\tau, Nz) \right] = \left[\Phi^{(N) '}(\tau, Nz) \right] = y^{-\frac{N}{2}}\, \sum_{j=0}^N\, y^j \equiv y^{-\frac{N}{2}} \frac{1-y^{N+1}}{1-y}.
$$
This means 
$$
\lim_{\tau\rightarrow i\infty}\, F(z_{0,b}) = 1 , ~~~ (\any b=1, \ldots, N),
$$
and thus we obtain 
$$
\lim_{\tau \rightarrow i \infty} \left| R_{0,b}(\tau) \right| =0, ~~~ (\any b=1, \ldots, N).
$$


~

\item[(ii) $a=1,\ldots, N-1, ~ b =0, \ldots, N $ :]

~

In this case, we first note
\begin{equation}
z_{a,b} = \xi_{a,b} + \tz_{a,b}, ~~~ \xi_{a,b} := \frac{a\tau+b}{N}, ~~~ \tz_{a,b} := - \frac{a\tau+b}{N(N+1)},
\label{decomp z ab}
\end{equation}
and the term $\xi_{a,b}$ can be interpreted as the spectral flow caused by $s_{(a,b)}$.
Therefore, it is enough to compare the behaviors of $\Phi^{(N)}_{(a,b)}(\tau, Nz)$ and   $\Phi^{(N)'}_{(a,b)}(\tau, Nz)$ around 
$z = \tz_{a,b}$ in place of examining $F(\tau, z)$ around $z= z_{a,b}$.

Recalling
$$
\Phi_{(a,b)}^{(N)}(\tau, Nz) = y^a \frac{\theta_1\left(\tau, (N+1)z + \xi_{a,b} \right)}{\theta_1 \left(\tau,z+ \xi_{a,b} \right)} , ~~~ (a\neq 0)
$$
we obtain
\begin{equation}
\Phi_{(a,b)}^{(N)}(\tau, Nz) \sim (z-\tz_{a,b}) \, q^{-\frac{a^2}{N(N+1)}+ \frac{a}{2(N+1)}}, ~~~ (z \sim \tz_{a,b}, ~~ \tau \sim i\infty).
\label{Phi ab behavior tz ab}
\end{equation}

On the other hand, since $\left[\Phi^{(N) '}_{(a,b)}(\tau, Nz) \right] = y^{-\frac{N}{2} + a}$, $(a\neq 0)$ holds, 
we also obtain 
\begin{equation}
\Phi_{(a,b)}^{ (N) '}(\tau, Nz) \sim \left(e^{2\pi i \tz_{a,b}}\right)^{-\frac{N}{2} +a}
\sim  q^{-\frac{a^2}{N(N+1)}+ \frac{a}{2(N+1)}}, ~~~ (z \sim \tz_{a,b}, ~~ \tau \sim i\infty).
\label{Phi' ab behavior tz ab}
\end{equation}
We thus conclude 
$$
\lim_{\tau \rightarrow  i \infty} \left| R_{a,b}(\tau) \right| < \infty, ~~~  (a=1,\ldots, N-1, ~ b =0, \ldots, N).
$$


~

\item[(iii) $a=N, ~b=0,\ldots , N$ : ]

~

Around $z= z_{N,b}$, we find 
$\th_1(\tau, (N+1)z) \sim \left(z-z_{N,b}\right) \, q^{-\frac{N^2}{2} + \frac{1}{8}}$ and 
$\th_1(\tau, z) \sim  q^{- \frac{N}{2(N+1)} + \frac{1}{8}}$,
and hence 
\begin{equation}
\Phi^{(N)}(\tau, Nz) \sim (z- z_{N,b}) \, q^{-\frac{N^2}{2}+ \frac{N}{2(N+1)}}, ~~~ (z \sim z_{N,b}, ~~ \tau \sim i\infty).
\label{Phi behavior z Nb}
\end{equation}

On the other hand, by using the decomposition \eqn{decomp z ab},  
we can evaluate $\Phi^{(N)'}(\tau, Nz)$ around $z = z_{N,b}$ as follows;
\begin{eqnarray}
\Phi^{(N) '}(\tau, Nz) 
&= & \Phi^{ (N) '}(\tau, N(\xi_{N,b} + \tz_{N,b})) 
\nn
&\sim&
q^{-\frac{1}{2} N(N+2)} e^{-2\pi i N(N+2) \tz_{N,b}} \times \left[\Phi^{(N) '}(\tau, N\tz_{N,b})\right]
\nn
&\sim & q^{-\frac{1}{2} N(N+2)} \left(e^{2\pi i \tz_{N,b}}\right)^{-N(N+2)+ \frac{N}{2}}
\nn
&\sim&  q^{-\frac{N^2}{2}+ \frac{N}{2(N+1)}}, ~~~ (z \sim z_{N,b}, ~~ \tau \sim i\infty).
\label{Phi' behavior z Nb}
\end{eqnarray}

Consequently, we again obtain 
$$
\lim_{\tau \rightarrow  i \infty} \left| R_{N,b}(\tau) \right| < \infty, ~~~  (b =0, \ldots, N).
$$

\end{description}

~

In this way, we have shown that 
the residue function $R_{a,b}(\tau)$ is finite at the cusp $\tau = i\infty$ for every simple pole $z=z_{a,b}$.



\newpage
\begin{thebibliography}{99}







\bibitem{ES-NH}
  T.~Eguchi and Y.~Sugawara,
  JHEP {\bf 1103}, 107 (2011)
  [arXiv:1012.5721 [hep-th]].
  
  
  
\bibitem{N=2 minimal}
  A.~B.~Zamolodchikov and V.~A.~Fateev,
  Sov.\ Phys.\ JETP {\bf 63}, 913 (1986)
  [Zh.\ Eksp.\ Teor.\ Fiz.\  {\bf 90}, 1553 (1986)];
 W.~Boucher, D.~Friedan and A.~Kent,
  Phys.\ Lett.\ B {\bf 172}, 316 (1986);
 P.~Di Vecchia, J.~L.~Petersen, M.~Yu and H.~B.~Zheng,
  Phys.\ Lett.\ B {\bf 174}, 280 (1986);
 F.~Ravanini and S.~-K.~Yang,
  Phys.\ Lett.\ B {\bf 195}, 202 (1987);
Z.~-a.~Qiu,
  Phys.\ Lett.\ B {\bf 188}, 207 (1987);
 Z.~-a.~Qiu,
  Phys.\ Lett.\ B {\bf 198}, 497 (1987).



  
  
  



\bibitem{KS}
Y.~Kazama and H.~Suzuki,
Nucl.\ Phys.\ B {\bf 321}, 232 (1989).


\bibitem{N=2LG} 
  E.~J.~Martinec,
  Phys.\ Lett.\ B {\bf 217}, 431 (1989);
  C.~Vafa and N.~P.~Warner,
  Phys.\ Lett.\ B {\bf 218}, 51 (1989);
  W.~Lerche, C.~Vafa and N.~P.~Warner,
  Nucl.\ Phys.\ B {\bf 324}, 427 (1989).



\bibitem{Witten-EG minimal} 
  E.~Witten,
  Int.\ J.\ Mod.\ Phys.\ A {\bf 9}, 4783 (1994)
  [hep-th/9304026].




\bibitem{EZ}
M. Eichler and D. Zagier, 
{\it``The Theory of Jacobi Forms,''}
(Birkh\"{a}user, 1985).




\bibitem{Troost}
  J.~Troost,
  JHEP {\bf 1006}, 104 (2010)
  [arXiv:1004.3649 [hep-th]].




\bibitem{ES-L}
T.~Eguchi and Y.~Sugawara,
JHEP {\bf 0401}, 025 (2004)
[arXiv:hep-th/0311141].



\bibitem{ES-BH}
  T.~Eguchi and Y.~Sugawara,
  JHEP {\bf 0405}, 014 (2004)
  [arXiv:hep-th/0403193].



\bibitem{ES-C}
  T.~Eguchi and Y.~Sugawara,
  JHEP {\bf 0501}, 027 (2005)
  [arXiv:hep-th/0411041].


\bibitem{Pol}
A. ~Polishchuk,
arXiv:math.AG/9810084.

\bibitem{STT}
A.~M.~Semikhatov, A.~Taormina and I.~Y.~Tipunin,
arXiv:math.qa/0311314.





\bibitem{Zwegers}
S. Zwegers, PhD thesis, 
``Mock Theta functions'', Utrecht University, 2002.






\bibitem{2DBH}
E. Witten, 
Phys. Rev. {\bf D44} (1991) 314;
G. Mandal, A. Sengupta and S. Wadia, 
Mod. Phys. Lett. {\bf A6} (1991) 1685;
I. Bars and D. Nemeschansky, 
Nucl. Phys. {\bf B348} (1991) 89;
S. Elizur, A. Forge and E. Rabinovici, 
Nucl. Phys. {\bf B359} (1991) 581;
R.~Dijkgraaf, H.~Verlinde and E.~Verlinde,
Nucl.\ Phys.\ B {\bf 371}, 269 (1992).






\bibitem{orb-ncpart}
  Y.~Sugawara,
JHEP {\bf 1201}, 098 (2012)  [arXiv:1109.3365 [hep-th]].  






\bibitem{IKPT}
  D.~Israel, C.~Kounnas, A.~Pakman and J.~Troost,
  JHEP {\bf 0406}, 033 (2004)
  [arXiv:hep-th/0403237].



\bibitem{HPT}
A.~Hanany, N.~Prezas and J.~Troost,
JHEP {\bf 0204}, 014 (2002)
[arXiv:hep-th/0202129].




\bibitem{LST}
A.~Giveon and D.~Kutasov,
JHEP {\bf 9910}, 034 (1999)
[arXiv:hep-th/9909110];
A.~Giveon and D.~Kutasov,
JHEP {\bf 0001}, 023 (2000)
[arXiv:hep-th/9911039].




\bibitem{Sug-NENS5} 
  Y.~Sugawara,
  JHEP {\bf 1210}, 159 (2012)
  [arXiv:1208.3534 [hep-th]].







\end{thebibliography}
\end{document}